%% file: Thesis-summary.tex
\documentclass[11pt]{article}

\usepackage[cp1251]{inputenc} 

\usepackage{amsmath}
\usepackage{amssymb}
\usepackage{amsthm}
\usepackage{latexsym}
\usepackage[dvips]{graphicx,color}
\usepackage{float}
\usepackage{array}
\usepackage{xcolor}
\usepackage{enumerate} 
\usepackage[shortlabels]{enumitem}

\usepackage{multirow}
\usepackage[hyphens]{url}
\usepackage{hyperref}
\usepackage{etoolbox}

\usepackage[square,sort,comma,numbers]{natbib}
\theoremstyle{plain}
\newtheorem{theorem}{Theorem}[section]

\newtheorem{prop}[theorem]{Proposition}
\newtheorem{corollary}[theorem]{Corollary}

\newtheorem{lemma}[theorem]{Lemma}
\newtheorem{hypothesis}[theorem]{Hypothesis}

\theoremstyle{definition}
\newtheorem{defn}[theorem]{Definition}

\theoremstyle{remark}
\newtheorem{remark}[theorem]{Remark}

\input{2018-macros}

\begin{document}

\title{\LARGE{A Generalization of the {\lt} Preservation Theorem}\\\vspace{4pt}\large{(Summary of dissertation including extensions of dissertation results)}}
\author{Abhisekh Sankaran\footnote{Current affiliation: Department of Computer Science and Technology, University of Cambridge, UK.}\\\\
Advisors: Supratik Chakraborty and Bharat Adsul}
\date{Indian Institute of Technology Bombay, Maharashtra, India.}

\maketitle

\begin{abstract}
This article gives a summary of the author's
Ph.D. dissertation~\cite{abhisekh-thesis}. In addition to an overview
of notions and results, it also provides sketches of various proofs
and simplified presentations of certain abstract results of the
dissertation, that concern tree representations of structures.
Further, some extensions of the dissertation results are
presented. These include the connections of the model-theoretic
notions introduced in the thesis with fixed parameter tractability and
notions in the structure theory of sparse graph classes. The
constructive aspects of the proofs of the model-theoretic results of
the dissertation are used to obtain (algorithmic) meta-kernels for
various dense graphs such as graphs of bounded clique-width and
subclasses of these like $m$-partite cographs and graph classes of
bounded shrub-depth. Finally, the article presents updated definitions
and results concerning logical fractals introduced
in~\cite{abhisekh-csl17} as a generalization of the Equivalent Bounded
Substructure Property from the dissertation. In particular, our
results show that (natural finitary adaptations of) both the upward
and downward versions of the L\"owenheim-Skolem theorem from classical
model theory can be recovered in a variety of algorithmically
interesting settings, and further in most cases, in effective form and
even for logics beyond first order logic.
\end{abstract}

\input{intro-3.0}

\input{overview-2.0}
\input{results-in-CMT}

\input{results-in-FMT}

\input{conclusion}

%%%%%%%%%%%%%%%%%%%%%%%% Bibliography %%%%%%%%%%%%%%%%%%%%%%%%%%%%%%

\bibliographystyle{plain}

\bibliography{Bibfiles/model-theory,Bibfiles/finite-model-theory,Bibfiles/databases-and-verification,Bibfiles/automata,Bibfiles/logic-over-special-graph-classes,Bibfiles/structural-graph-theory,Bibfiles/algorithms,Bibfiles/misc,Bibfiles/self}

%%%%%%%%%%%%%%%%%%%%% ---------------- %%%%%%%%%%%%%%%%%%%%%%%%%%%%%

%%%%%%%%%%%%%%%%%%%%%%%%%% Appendices %%%%%%%%%%%%%%%%%%%%%%%%%%%%%%

\input{appendix}

%%%%%%%%%%%%%%%%%%%%%%% ---------------- %%%%%%%%%%%%%%%%%%%%%%%%%%%

\end{document}

%% file: 2018-macros.tex
\newcommand{\tbf}[1]{\textbf{#1}}
\newcommand{\mc}[1]{\mathcal{#1}}

\newcommand{\mf}[1]{\mathfrak{#1}}

\newcommand{\mequiv}[1]{\ensuremath{\equiv_{#1, \mso}}}
\newcommand{\lequiv}[1]{\ensuremath{\equiv_{#1, \mc{L}}}}

\newcommand{\fo}{\ensuremath{\text{FO}}}
\newcommand{\mso}{\ensuremath{\text{MSO}}}

\newcommand{\hpc}[1]{\ensuremath{h\text{-}PC({#1})}}

\newcommand{\lt}{{{\L}o{\'s}-Tarski}}

\newcommand{\glt}[1]{\ensuremath{\mathsf{GLT}({#1})}}
\newcommand{\lglt}[1]{\ensuremath{\mc{L}\text{-}\mathsf{GLT}({#1})}}

\newcommand{\hpt}{\ensuremath{\mathsf{HPT}}}

\newcommand{\lghpt}[1]{\ensuremath{\mc{L}\text{-}\mathsf{GHPT}({#1})}}

\newcommand{\lebsp}[1]{\ensuremath{\mc{L}\text{-}\mathsf{EBSP}({#1})}}
\newcommand{\febsp}[1]{\ensuremath{\fo\text{-}\mathsf{EBSP}({#1})}}
\newcommand{\mebsp}[1]{\ensuremath{\mso\text{-}\mathsf{EBSP}({#1})}}

\newcommand{\reflebsp}{\ensuremath{\mc{L}\text{-}\mathsf{EBSP}}}

\newcommand{\hlebsp}[2]{\ensuremath{h\text{-}\mc{L}\text{-}\mathsf{EBSP}({#1},
    {#2})}}

\newcommand{\refdls}{\ensuremath{\mathsf{DLS}}}

\newcommand{\dls}{downward L\"owenheim-Skolem}

\newcommand{\eat}[1]{}

\newcommand{\tree}[1]{\ensuremath{\mathsf{#1}}}

\newcommand{\sigmaint}{\ensuremath{\Sigma_{\text{int}}}}
\newcommand{\sigmarank}{\ensuremath{\Sigma_{\text{rank}}}}
\newcommand{\sigmaleaf}{\ensuremath{{\Sigma}_{\text{leaf}}}}

\newcommand{\nesword}[1]{\ensuremath{\mathsf{#1}}}

\newcommand{\cl}[1]{\ensuremath{\mathcal{#1}}}

\newcommand{\str}[1]{\ensuremath{\mathsf{Str}(#1)}}

\newcommand{\scale}[1]{\langle #1 \rangle}

%% file: intro-3.0.tex
\section{Introduction}\label{section:introduction}

Classical model theory is a subject within mathematical logic that
studies the relationship between a formal language and its
interpretations also called structures or models~\cite{chang-keisler}.
Amongst the earliest areas of study in classical model theory, is a
class of results called \emph{preservation theorems}. A preservation
theorem syntactically characterizes classes of arbitrary structures
(structures that could be finite or infinite) that are closed under a
given model-theoretic operation.  For instance, the class of all
cliques is preserved under substructures (induced subgraphs in this
context). This class is also defined by the first order (FO) sentence
that says ``for all (vertices) $x$ and for all (vertices) $y$, (there
is an) edge between $x$ and $y$''. The latter is a ``universal''
sentence, i.e. an FO description that contains only universal
quantifications. One of the earliest preservation theorems of
classical model theory, the {\lt} theorem, proven by Jerzy {\L}o{\'s}
and Alfred Tarski in 1954-55, says that universal sentences are
indeed \emph{expressively complete} for preservation under
substructures.

Technically, the {\lt} theorem states that a class of arbitrary
structures defined by an FO sentence is preserved under substructures
if, and only if, it is definable by a universal
sentence~\cite{chang-keisler}.  The theorem in ``dual'' form
characterizes extension closed FO definable classes of arbitrary
structures in terms of ``existential'' sentences.  Both forms of the
theorem extend to theories (sets of sentences) as well.  The {\lt}
theorem is historically important for classical model theory since its
proof constituted the earliest applications of the FO Compactness
theorem which is now regarded as one of the pillars of model
theory. Further, the proof triggered off an extensive study of
preservation theorems in which various other model-theoretic
operations like homomorphisms, unions of chains, direct products,
etc. were taken up and preservation theorems for these operations were
proven not just for FO but even extensions of it, like infinitary
logics~\cite{hodges-history}. These investigations contributed
significantly to the development of the (then young) subject of
classical model theory.

In 1973, Fagin proved a landmark result characterizing the complexity
class \textsf{NP} in terms of existential second order logic.  This
result began the area of \emph{finite model theory}, whose aims are
similar to classical model theory, namely the study of the expressive
power of formal languages, but now the structures under consideration
are only finite. Finite model theory~\cite{libkin} is closely
connected with computer science since many disciplines within the
latter use formal languages, such as programming languages, database
query languages or specification languages, and further the structures
arising in these disciplines are often finite.  It is natural to ask
if the results and techniques of classical model theory can be carried
over to the finite too. It turns out that the class of all finite
structures is very poorly behaved from the model-theoretic
perspective. The {\lt} theorem fails, the L\"owenheim-Skolem theorem
(a central result in model theory) becomes meaningless, and important
concepts like saturation and homogeneity that provide elegant results
in the infinite, trivialize in the finite~\cite{rosen}. A few theorems
survive, such as the homomorphism preservation
theorem~\cite{rossman-hom} and van Benthem's modal characterization
theorem~\cite{rosen-van-benthem}, but these are rare. This inspired
the research programme of identifying classes of finite structures
over which results of classical model theory could be recovered. The
pioneering steps were taken in~\cite{dawar-hom, dawar-pres-under-ext}
where it was shown that algorithmically interesting classes, in
particular sparse structures, such as those that are acyclic, of
bounded degree or of bounded tree-width, under reasonable closure
assumptions satisfy the (relativized versions of the) {\lt} and
homomorphism preservation theorems. (Note that the truth of a
preservation theorem over a class does not imply its truth over a
subclass; going to the subclass weakens both sides of the equivalence
given by the theorem). The homomorphism preservation theorem was later
shown to be true over hereditary quasi-wide classes too (under mild
assumptions)~\cite{dawar-quasi-wide}. Quasi-wide structures are
exactly nowhere dense structures under
hereditariness~\cite{nesetril-sparsity}, and the latter are regarded
as the natural limit of algorithmic techniques for sparse
classes~\cite{grohe-nowhere-dense}.

While a preservation theorem provides a syntactic characterization of
a preservation property, the same theorem flipped around, can also be
seen as providing a semantic characterization (via a preservation
property) of a syntactic class of sentences. Thus the {\lt} theorem
characterizes universal and existential sentences in terms of
preservation under substructures and extensions, respectively. It is
well known that if the vocabulary contains only relation symbols, then
the sizes of the minimal models of a sentence preserved under
extensions, are no larger than the number of quantifiers in any
equivalent existential sentence~\cite{dawar-pres-under-ext}. Thus the
{\lt} theorem, besides giving a syntax-semantics correspondence, also
yields a relation between a quantitative model-theoretic property
(size of minimal models) of a sentence in a semantic class and the
count of quantifiers in an equivalent sentence in the corresponding
syntactic class.  Two subclasses of FO that are semantically richer
than the universal and existential classes of sentences, are the
$\Sigma^0_2$ and $\Pi^0_2$ classes of prenex FO sentences having two
blocks of quantifiers, with the leading block being existential and
universal respectively. These subclasses of FO arise in a wide variety
of areas in computer science: decision procedures for SAT, program
verification, SMT solvers and program synthesis (where $\Sigma^0_2$ is
called the \emph{Bernays-Sch\"onfinkel-Ramsey} class or
\emph{effectively propositional logic}), and databases, particularly
data exchange, data integration and query answering over RDF and OWL
knowledge (where $\Pi^0_2$ is called the \emph{forall-existential}
fragment)~\cite{bjorner,fontaine-BSR,data-exch,mathew-2}. Further, a
number of fixed parameter tractable (FPT) problems, as well as
important notions from finite model theory and the structure theory of
sparse graphs, turn out to be naturally described in the $\Sigma^0_2$
and $\Pi^0_2$ fragments (see Appendix~\ref{appendix}).  The classical
model theory literature contains several semantic characterizations,
over arbitrary structures, for these syntactic fragments, in terms of
notions such as ascending chains, descending chains, and Keisler's
1-sandwiches~\cite{chang-keisler}.

However, there are two major drawbacks of the mentioned notions from
the perspective of computer science. First, none of these notions
enables relating \emph{quantifier counts} in the syntactic classes to
any model-theoretic properties. Second, all of these notions, being
infinitary, trivialize in the finite. With the result that there are
no preservation theorems in the finite, that semantically characterize
the $\Sigma^0_2$ and $\Pi^0_2$ classes, or their subclasses with given
(non-zero) quantifier counts. Next, we observe that within the
programme of recovering classical model theory in the finite, the
investigations so far have been only over \emph{sparse}
structures. After the momentous recent result characterizing nowhere
dense classes as the largest subgraph-closed classes of sparse graphs
admitting FO algorithmic meta-theorems~\cite{grohe-nowhere-dense}, the
focus has shifted to \emph{dense} structures, such as posets,
subclasses of bounded clique-width graphs, and structures that are
first-order interpretable in sparse classes, for algorithmic
meta-theorems and structural
studies~\cite{bounded-width,shrub-depth-FO-equals-MSO,dense-classes-first-paper,FO-interpretation-bounded-expansion}. However,
there has been no research on the model-theoretic properties of these
classes, akin to that done for sparse structures. We observe also that
the studies for sparse structures have considered only the {\lt} and
homomorphism preservation theorems from model theory.  This thesis is
motivated by all of the above issues, and takes the first steps
towards addressing them.

%% file: overview-2.0.tex
\section{Overview of the Thesis}\label{section:overview}

\tbf{New parameterized preservation properties}: The starting point of
the thesis~\cite{abhisekh-thesis} is the introduction of two dual
parameterized preservation properties that generalize the properties
of preservation under substructures and preservation under extensions
(Chapter 3). We call these respectively \emph{preservation under
  substructures modulo $k$-cruxes}, denoted $PSC(k)$, and
\emph{preservation under $k$-ary covered extensions}, denoted
$PCE(k)$, where $k$ is a natural number.

\begin{defn}[Defn. 3.1.1, Chp. 3]\label{defn:PSC(k)}
A sentence $\varphi$ is said to be $PSC(k)$ if for every model
$\mf{A}$ of $\varphi$, there is a subset $C$ of the universe of
$\mf{A}$, of size $\leq k$, such that if $\mf{B}$ is a substructure of
$\mf{A}$ and $\mf{B}$ contains $C$, then $\mf{B}$ is also a model of
$\varphi$.  We call $C$ a \emph{$k$-crux of $\mf{A}$ with respect to
  $\varphi$}.
\end{defn}

\begin{defn}[Defn. 3.2.1 \& Defn. 3.2.4, Chp. 3]\label{defn:k-ary-covered-ext}
Given a structure $\mf{A}$, a non-empty collection $\mc{R}$ of
substructures of $\mf{A}$ is said to be a \emph{$k$-ary cover} of
$\mf{A}$ if for every subset $C$ of the universe of $\mf{A}$, of size
$\leq k$, there is a structure in $\mc{R}$ that contains $C$.  We call
$\mf{A}$ a \emph{$k$-ary covered extension} of $\mc{R}$. We now say a
sentence $\varphi$ is $PCE(k)$ if for every collection $\mc{R}$ of
models of $\varphi$, if $\mf{A}$ is a $k$-ary covered extension of
$\mc{R}$, then $\mf{A}$ is also a model of $\varphi$.
\end{defn}

As an example, consider the sentence $\varphi$ over graphs that
asserts ``there is a dominating set of size $\leq k$''. In any model
of $\varphi$, any dominating set witnessing $\varphi$ is a $k$-crux of
the model with respect to $\varphi$; then $\varphi$ is $PSC(k)$. One
observes that $\neg \varphi$ is $PCE(k)$; indeed this an instance of
the more general observation that a sentence is $PSC(k)$ if, and only
if, its negation is $PCE(k)$. We see that $PSC(\cdot)$ and
$PCE(\cdot)$ form semantic hierarchies under inclusion, that strictly
increase with $k$ (the sentence that asserts ``there are $\ge k$
vertices in the graph'' is $PSC(k)$ but not $PSC(k-1)$), and whose $k
= 0$ levels correspond exactly to preservation under substructures and
extensions respectively.

The introduced properties turn out to be widely realized in computer
science (see Appendix~\ref{appendix}). The reason as it turns out, is
connected with the definability of these properties using syntactical
structure that entails the properties. Specifically, any $\exists^k
\forall^*$ sentence, i.e. a $\Sigma^0_2$ sentence having $k$
quantifiers in its leading block, is $PSC(k)$: the witnesses to the
existential quantifiers in any model form a $k$-crux in the model. For
instance, the dominating set example above is definable using the
$\exists^k \forall^*$ sentence $\varphi = \exists x_1 \ldots \exists
x_k$ $ \forall y \bigvee_{i = 1}^{i = k} (y = x_i \vee E(y,
x_i))$. Dually any $\forall^k \exists^*$ sentence is $PCE(k)$. The
natural question that arises is whether the mentioned syntactic
fragments characterize their corresponding preservation
properties. This question is the central thrust of this thesis, and
indeed we show that in a number of scenarios, infinitary and finitary, the
question has an affirmative answer via the following preservation
theorem, that we call the \emph{generalized {\lt}} theorem, or
$\glt{k}$ in short: an FO sentence is $PSC(k)$, respectively $PCE(k)$,
if, and only if, it is equivalent to an $\exists^k \forall^*$
sentence, respectively a $\forall^k \exists^*$ sentence. The case of
$k = 0$ is verily the {\lt} theorem.

\vspace{4pt} \noindent \tbf{Classical model theory results:} In Part I
of the thesis, we investigate $\glt{k}$, its variants and extensions,
over arbitrary structures. Specifically, we first show (in Chapter
4.1) that $\glt{k}$ holds over any elementary class of arbitrary
structures (i.e. a class definable using an FO theory).  To the best
of our knowledge, $\glt{k}$ is the first in the classical model theory
literature to relate natural quantitative properties of models of
sentences in a semantic class, to counts of leading quantifiers in
equivalent $\Sigma^0_2$ or $\Pi^0_2$ sentences. It thereby provides
new and finer characterizations of these syntactic classes than those
in the literature, and moreover, via notions that are combinatorial in
nature. Precisely due to the latter, the notions remain non-trivial in
the finite too, making it meaningful to ask which classes of finite
structures satisfy $\glt{k}$.

We continue our investigations over arbitrary structures, generalizing
$\glt{k}$ by extending the definitions of $PSC(k)$ and $PCE(k)$ in two
different directions: one, by considering cruxes of sizes, and covers
of arities, less than an infinite cardinal $\lambda$ (Chapter 4.2),
and the other, by considering theories instead of sentences (Chapter
5). For the former, we show that our ``infinitary'' properties
$PSC(\lambda)$ and $PCE(\lambda)$ characterize all of $\Sigma^0_2$ and
$\Pi^0_2$ respectively for sentences, and thus coincide with their
finitary counterparts taken together over all $k$. This is indeed a
Compactness-like phenomenon which we employ as a new technique to
(re)prove well-known inexpressibility results of FO, such as
acyclicity, bipartiteness, etc.~\cite{inexp-TR} The other
aforementioned direction, namely the extension of $\glt{k}$ to
theories, yields us results that differ from the ``sentential''
results for $\glt{k}$ in many important ways. Specifically, while
$PCE(k)$ theories do characterize theories of $\forall^k \exists^*$
sentences, $PSC(k)$ theories fall way short of doing analogously for
theories of $\exists^k \forall^*$ sentences. Indeed the $\exists
\forall^*$ theory (just one existential quantifier) defining the class
of (arbitrary) graphs not containing for any $n$, a vertex cycle cover
consisting of cycles of length $\leq n$, is not even $PSC(\aleph_0)$
(and hence not $PSC(k)$ for any $k$). The infinitary properties are
more well-behaved in contrast: for $\lambda \ge \aleph_0$,
$PCE(\lambda)$ characterizes $\Pi^0_2$ theories, while for $\lambda
\ge \aleph_1$, $PSC(\lambda)$ characterizes $\Sigma^0_2$ theories, so
instead of an ``infinite to finite collapse'' of the $PSC$ and $PCE$
hierarchies as in the case of sentences, for theories there is a such
collapse in the infinite realm down to the first or second level.

The above investigations leave open the characterization of $PSC(k)$
theories. While this question has still evaded our best efforts to
answer it, we do give partial results which constitute the most
technically involved parts of Part I of the thesis. Specifically, we
first show that a $PSC(k)$ theory is always equivalent to a
$\Sigma^0_2$ theory, and then using key insights gained from its
proof, we refine this result by showing that under a well-motivated
model-theoretic hypothesis, a $PSC(k)$ theory is always equivalent to
an $\exists^k \forall^*$ theory. The proof of the latter introduces a
novel technique of characterizing semantically defined FO theories by
\emph{going outside FO.}  In particular, we show under the mentioned
hypothesis, a characterization of $PSC(k)$ theories in terms of
sentences of an \emph{infinitary logic}, and then ``compile'' the
infinitary sentences down to their equivalent syntactically defined FO
theories, by using suitable ``finite approximations'' of the
former. We believe that this technique of characterizing FO theories
might be of independent interest. We conclude Part I of the thesis on
this note (Chapter 6), raising various natural questions for future
work (over arbitrary structures), notable amongst them being a
characterization of $PSC(k)$ theories, and a lifting of our results to
characterize prefix subclasses of FO having more quantifier
alternations than just one.

\vspace{4pt} \noindent \tbf{Finite model theory results:} In Part II
of the thesis, we turn our attention to finite structures to
investigate for $\glt{k}$.  The failure of the {\lt} theorem over the
class of all finite structures shows that $\glt{0}$ fails over this
class. We show a stronger failure: $\glt{k}$ fails for each $k$
(Chapter 8). Indeed for each $k$, we construct a sentence $\varphi_k$
that is hereditary (preserved under substructures) over the class of
all finite structures (so $\varphi_k$ is also $PSC(k)$ over this
class), but that is not equivalent over this class to any $\exists^k
\forall^*$ sentence. This strengthens the known failure of the {\lt}
theorem in the finite, by showing that not just universal sentences,
but even $\exists^k \forall^*$ sentences for each $k$, fail to capture
hereditary FO properties over all finite structures. Investigating
further, we also show that the classes of sparse graphs mentioned
earlier that admit the {\lt} theorem (graphs that are acyclic, of
bounded degree or of bounded tree-width), do not admit $\glt{k}$ for
any $k \ge 2$. This is because for $k \ge 2$, (the satisfaction of)
$\glt{k}$ forces a (hereditary) class to have bounded induced path
lengths. (It is interesting to note here that a \emph{preservation
  theorem} enforces a \emph{structural condition}.)  We therefore
consider classes of \emph{dense structures}, in particular those of
significant current interest from the perspectives of algorithmic
meta-theorems and structural graph theory. Remarkably, these turn out
to not just satisfy $\glt{k}$, but various other classical model
theory results as well, none of which were earlier known to hold over
them.

We do all our investigations within an abstract framework that
incorporates crucial observations about the structural and logical
properties of the dense classes referred to. Specifically, the
framework considers classes of structures that admit, what we call,
\emph{$\mc{L}$-good tree representations}, where $\mc{L}$ is either
FO, or monadic second order logic (MSO). An $\mc{L}$-good tree
representation is a tree whose leaf nodes are labeled with simple
structures (typically singletons) and whose internal nodes are labeled
with operations on structures, that satisfy a few natural monotonicity
properties with respect to isomorphic embedding, along with the
\emph{Feferman-Vaught composition (FVC)} property. An operation has
the $\mc{L}$-FVC property if for each $m$, the $\mc{L}[m]$ theory of
the operation's output is \emph{determined} by the $\mc{L}[m]$
theories of the operation's inputs. Here, the $\mc{L}[m]$ theory of a
structure is the set of $\mc{L}$ sentences of quantifier nesting depth
$m$ that are true in the structure.  As examples, disjoint union has
the $\mso$-FVC property, while Cartesian product has the $\fo$-FVC
property.

Towards our model-theoretic results, we first show for each $k$, that
if the class $\cl{S}_k$ of $k$-labeled (i.e. labeled with
$\{1, \ldots, k\}$, partially) versions of the structures of a given
class $\cl{S}$ admits $\mc{L}$-good tree representations, then
$\cl{S}_k$ satisfies a natural finitary analogue of the {\dls}
($\refdls$) property from classical model theory, and $\cl{S}$
satisfies a ``$k$-rooted'' version of this analogue.  We call the
latter version the \emph{$\mc{L}$-Equivalent Bounded Substructure
Property}, denoted $\lebsp{\cl{S}, k}$. Intuitively, $\lebsp{\cl{S},
k}$ asserts that any structure $\mf{A}$ in $\cl{S}$ contains a small
substructure $\mf{B}$ in $\cl{S}$, that is ``$\mc{L}[m]$-similar'' to
$\mf{A}$, in the sense that $\mf{A}$ and $\mf{B}$ have the same
$\mc{L}[m]$ theory. The bound on the size of $\mf{B}$ is a function of
$m$ alone (if $\cl{S}$ and $k$ are fixed); we call the latter a
``witness function''. Further, such a substructure $\mf{B}$ can always
be found ``around'' any given set of at most $k$ elements of
$\mf{A}$. Comparing this with the $\refdls$ property that states that
every infinite structure has a countable FO-similar substructure
``around'' any countable subset of the former, we see (right away)
that $\lebsp{\cl{S}, k}$ can be regarded as a finitary analogue of the
$\refdls$. While $\lebsp{\cl{S}, k}$ puts no constraints on the
witness function, we show that a class $\cl{S}$ that admits
$\mc{L}$-good tree representations for $\cl{S}_k$, actually satisfies
$\lebsp{\cl{S}, k}$ with a \emph{computable} witness function.

We now go on to show that $\lebsp{\cl{S}, k}$ in turn always entails
(irrespective of the abstract framework), an ``$\mc{L}$-version'' of
$\glt{k}$, denoted $\lglt{k}$, which characterizes $\mc{L}$ sentences
that are $PSC(k)$ in terms of $\exists^k \forall^*$ FO sentences (thus
$\lglt{k}$ is stronger than $\glt{k}$ which is only for FO).
Furthermore, if $\lebsp{\cl{S}, k}$ holds with a computable witness
function, then $\lglt{k}$ holds in effective form, i.e. the
translation from an $\mc{L}$ sentence that is $PSC(k)$ to its
equivalent $\exists^k \forall^*$ sentence, is effective.  It turns out
that $\lebsp{\cl{S}, k}$ (with computable witness functions) also
entails (an effective version of) the homomorphism preservation
theorem ($\hpt$) and a parameterized generalization of it along the
lines of $\lglt{k}$.  This we prove by showing $\lebsp{\cl{S}, k}$ to
entail a ``homomorphic'' variant of itself, which in turn entails the
said generalization of the $\hpt$. Summing up our analysis above then,
we get that any class of structures falling within our abstract
framework is model-theoretically very well behaved: it admits (i) the
$\mc{L}$-FVC property for operations that construct its structures,
(ii) two variants of a finitary and ``computable'' analogue of the
$\refdls$, a ``substructure'' variant and a ``homomorphic'' variant,
and finally, effective $\mc{L}$-versions of (iii) the {\lt} theorem
and its parameterized generalization $\glt{k}$, and (iv) the $\hpt$
and its parameterized generalization akin to $\glt{k}$. (Chapters
9, 10.1 and 11.3 contain the above results.)

We now show that various interesting dense classes of structures come
within the fold of our abstract framework, by proving the relevant FVC
theorems for these classes (Chapters 10.2, 10.3 and 10.4.). The dense
classes we investigate are broadly of two kinds: special types of
labeled posets and special classes of graphs. For the former, we show
our results for words, trees (unordered, ordered, ranked, partially
ranked) and nested words over a finite alphabet, and further regular
subclasses of these. For the latter, we consider hereditary subclasses
of graphs of bounded clique-width whose $k$-expressions exclude
relabeling, such as threshold graphs, cographs, graph classes of
bounded shrub-depth (which include graphs of bounded tree-depth),
$m$-partite threshold graphs, and $m$-partite
cographs~\cite{shrub-depth}. These classes have attracted significant
current interest due to their excellent algorithmic and logical
properties: for instance, bounded shrub-depth classes admit
algorithmic meta-theorems for CMSO (a counting extension of MSO) with
elementary parameter dependence; also $\mso$ and $\fo$ have equal
expressive powers over these classes~\cite{shrub-depth-FO-equals-MSO}.
We go further to show a number of closure properties of $\lebsp{\cdot,
  \cdot}$, such as closure under various set-theoretic operations and
logical interpretations, enabling us to construct several
model-theoretically well-behaved classes from ones that are already so
(preserving the computational aspects of the model-theoretic results).

In addition to the classes described above, we identify another
important collection of classes that admit our model-theoretic
results: classes well-quasi-ordered (w.q.o.) under isomorphic
embedding. Well-quasi-ordering is a concept that has time and again
proved to be of importance for computer
science~\cite{dagstuhl-seminar-report}. We first observe that for any
given class, w.q.o. under isomorphic embedding is equivalent to the
class satisfying the following ``unconditional version'' of the {\lt}
theorem: \emph{any} isomorphism-closed hereditary subclass (not
necessarily apriori known to be definable in a logic) is always
defined by a universal FO sentence that asserts the exclusion under
isomorphic embedding, of a finite set of structures from the class
(the so called ``forbidden'' set). We now show (Chapter 11.2) that if
a class $\cl{S}$ is such that $\cl{S}_k$ (its $k$-labeled version) is
w.q.o. under isomorphic embedding, then $\cl{S}$ satisfies
$\lebsp{\cl{S}, k}$, and hence all other model-theoretic results the
latter entails. The witness functions need not be computable though,
so that the preservation theorems entailed by $\lebsp{\cl{S}, k}$ need
not hold in effective form. In recent years, the study of graph
classes w.q.o./''labeled-w.q.o.'' under induced subgraphs has become
an active research theme, particularly in the context of analyzing
their clique-width. Cographs, $k$-letter graphs and $k$-uniform graphs
are examples of (dense) graph classes that are w.q.o. under induced
subgraphs (in fact, labeled-w.q.o. as
well)~\cite{wqo-vs-clique-width}. While all these graph classes have
bounded clique-width, it was shown in~\cite{wqo-vs-clique-width} that
there are hereditary graph classes, so called ``power graphs'', of
bounded induced path lengths and of unbounded clique-width (and hence
dense) that are w.q.o. under induced subgraphs. This then adds to our
list of model-theoretically well-behaved classes, various dense graph
classes, including those of unbounded clique-width as well.

\vspace{4pt} \noindent \tbf{Extensions of thesis results:} While
preparing for his Ph.D. defence, the author realized that the proof
establishing $\lebsp{\cdot, \cdot}$ within the abstract framework, is
really constructive in nature. In fact, the small $\mc{L}[m]$-similar
substructure for any structure as given by $\lebsp{\cdot, \cdot}$, is
actually computable in time \emph{linear} in the size of an
$\mc{L}$-good tree representation of the structure. Observing that the
mentioned substructure is indeed a ``meta-kernel'', we get linear time
algorithmic meta-theorems for FO/MSO model checking, for any class
that falls within the framework. (These results were subsequently
published in~\cite{abhisekh-csl17}.) Going further, it also dawned
that the scope of the abstract framework could be widened by not
insisting on isomorphic embedding in the monotonicity conditions, but
rather allowing for other wider relations as well. This relaxation
brings many more interesting classes into the ambit of the framework,
including sparse graphs: for instance, the class of all trees is
admitted by considering the homomorphic image relation, and more
generally, bounded tree-width graphs are admitted by considering the
minor relation.

Going still further, it was perceived that $\lebsp{\cdot, \cdot}$
really asserts ``logical self-similarity'' ``under substructure'' at
``small scales''. This suggested a strengthening of $\lebsp{\cdot,
  \cdot}$ that asserts logical self-similarity under any given
relation ``at all scales'' for a suitable notion of scale, as a
logical adaptation of the extensively studied \emph{fractal} property
from mathematics. We call this the \emph{logical fractal} property
``under'' the given relation. As shown in
Section~\ref{section:post-thesis}, this property turns out to be
ubiquitous in computer science. Further, for any class falling inside
the abstract framework relaxed to a relation $\preceq$, there is an
FPT algorithm parameterized by the quantifier nesting depth $m$ of the
logic $\mc{L}$ that, given a structure from the class, produces an
$\mc{L}[m]$-similar $\preceq$-related structure at \emph{any given
  scale}, in linear time. This yields the aforementioned
meta-kernelization as a special case. Observing that the logical
fractal property is a finitary adaptation of the ``full''
L\"owenheim-Skolem theorem proven by Mal'tsev in
1936~\cite{chang-keisler}, our generalizations above post the thesis
submission, based on the abstract framework and the proof ideas for it
as contained in the thesis, demonstrate that suitably adapted versions
of classical model theoretic results can be widespread in the finite.

\vspace{3pt} \noindent \tbf{Organization of the report.} In the
forthcoming sections, we give more technical details for the overview
presented above. In Sections~\ref{section:glt(k)-for-sentences}
and~\ref{section:glt-k-for-theories} we discuss $\glt{k}$ and its
variants for sentences and theories, over arbitrary structures.
Turning to the finite, we present in
Section~\ref{section:abstract-framework} the abstract framework within
which various model-theoretic results are established.  Applications
of the framework are discussed in Section~\ref{section:applications}
and connections with well-quasi-ordering in Section~\ref{section:wqo}.
Logical fractals and algorithmic results are presented in
Section~\ref{section:post-thesis}, before concluding the report in
Section~\ref{section:conclusion}. Finally, Appendix~\ref{appendix}
shows the relevance of our notions to computer science.

%% file: results-in-CMT.tex
\vspace{15pt}
\begin{center}
\LARGE{\textsc{Part I: Classical Model Theory}}\label{section:intro-results-in-CMT}
\end{center}
\vspace{2pt}
\input{glt-k-for-sentences-3.0}

\input{glt-k-for-theories-3.0}

%% file: glt-k-for-sentences-3.0.tex
\section{The Generalized {\lt} Theorem for Sentences}\label{section:glt(k)-for-sentences}

We syntactically characterize our preservation properties via the
following \emph{generalized {\lt} theorem}, abbreviated $\glt{k}$. The
{\lt} theorem is exactly $\glt{k}$ when $k$ equals $0$.

\begin{theorem}[$\glt{k}$; Thm. 4.1.1, Chp. 4]\label{theorem:glt(k)}
The following are true for each $k \in \mathbb{N}$.
\begin{enumerate}[nosep]
\item A sentence $\varphi$ is $PSC(k)$ if, and only if, $\varphi$ is
  equivalent to a $\Sigma^0_2$ sentence that has $k$ existential
  quantifiers.\label{theorem:glt(k)-subst}
\item A sentence $\varphi$ is $PCE(k)$ if, and only if, $\varphi$ is
  equivalent to a $\Pi^0_2$ sentence that has $k$ universal
  quantifiers.\label{theorem:glt(k)-ext}
\end{enumerate}
\end{theorem}

\begin{corollary}[Cor. 4.1.2, Chp. 4]\label{corollary:glt-PSC-PCE}
Let $PSC \Leftrightarrow \bigvee_{k \ge 0} PSC(k)$ and $PCE
\Leftrightarrow \bigvee_{k \ge 0} PCE(k)$. %The following are true.
\begin{enumerate}[nosep]
\item A sentence $\varphi$ is $PSC$ if, and only if, $\varphi$ is
  equivalent to a $\Sigma^0_2$ sentence.
\item A sentence $\varphi$ is $PCE$ if, and only if, $\varphi$ is
  equivalent to a $\Pi^0_2$ sentence.
\end{enumerate}
\end{corollary}

We give two proofs of Theorem~\ref{theorem:glt(k)}, one based on
saturated structures and the other based on ascending chains of
structures. Both proofs first show the characterization for $PCE(k)$,
and then ``dualize'' it to obtain the characterization for $PSC(k)$.
The non-trivial direction is the semantics-implies-syntax direction,
and in both proofs, we show that a $PCE(k)$ sentence $\varphi$ is
equivalent to the set $\Gamma$ of $\forall^k \exists^*$ sentences that
are entailed by it; then one application of Compactness yields a
finite subset of $\Gamma$, and hence a single $\forall^k \exists^*$
sentence, that is equivalent to $\varphi$.  We sketch below the
saturation based method of showing the equivalence of the mentioned
$\varphi$ and $\Gamma$.

We first show for any fixed infinite cardinal $\lambda$, that $\varphi
\leftrightarrow \Gamma$ holds over the class of
\emph{$\lambda$-saturated} structures. These structures possess
several nice model-theoretic features; of relevance to us is the
feature by which such a structure contains (as substructure) an
isomorphic copy of every structure of size $\leq \lambda$, that
realizes the universal type (i.e. the set of universal formulae that
are true) of a finite tuple of elements of the $\lambda$-saturated
structure. We now show that the universal type of any $k$-tuple
$\bar{a}$ of a $\lambda$-saturated model $\mf{A}$ of $\Gamma$, is
realized in a model $\mf{B}_{\bar{a}}$ of $\varphi$, of size $\leq
\lambda$; the latter then embeds into $\mf{A}$, further ``around''
$\bar{a}$, due to the $\lambda$-saturatedness of $\mf{A}$. The
collection of all such $\mf{B}_{\bar{a}}$'s forms a $k$-ary cover of
$\mf{A}$, making $\mf{A}$ a model of $\varphi$ as the latter is
$PCE(k)$.  We complete the proof by transferring the truth of $\varphi
\leftrightarrow \Gamma$ over $\lambda$-saturated structures, to over
all structures, using the fact that any (arbitrary) structure has an
elementary extension that is $\lambda$-saturated for some $\lambda$.

\vspace{3pt} \noindent \tbf{``Infinitary'' variants of $PSC(k)$ and
  $PCE(k)$}: The notions of $PSC(k)$ and $PCE(k)$ can be naturally
generalized to their infinitary counterparts obtained by respectively
allowing cruxes to be of sizes $< \lambda$ and covers to be of arities
$< \lambda$, for an infinite cardinal $\lambda$; we call these
variants $PSC(\lambda)$ and $PCE(\lambda)$ respectively. These
properties are dual in the same sense as $PSC(k)$ and $PCE(k)$
are. The following characterizations can be shown analogously to
Theorem~\ref{theorem:glt(k)}.

\begin{theorem}[Thm. 4.2.6, Chp. 4]
The following are true for each infinite cardinal $\lambda$.
\begin{enumerate}[nosep]
\item A sentence $\varphi$ is $PSC(\lambda)$ if, and only if,
  $\varphi$ is equivalent to a $\Sigma^0_2$ sentence.
\item A sentence $\varphi$ is $PCE(\lambda)$ if, and only if,
  $\varphi$ is equivalent to a $\Pi^0_2$ sentence.
\end{enumerate}
\end{theorem}

\begin{corollary}[Cor. 4.2.7, Chp. 4]\label{corollary:PSC-lambda-equals-PSC}
For each infinite cardinal $\lambda$, a sentence is $PSC(\lambda)$
(resp. $PCE(\lambda)$) if, and only if, it is $PSC(k)$
(resp. $PCE(k)$) for some $k \in \mathbb{N}$.
\end{corollary}

As mentioned in Section~\ref{section:overview}, the ``Compactness
flavour'' of the above corollary enables us to give a new technique,
via \emph{preservation properties}, to analyse the expressive power of
FO.  Corollary~\ref{corollary:PSC-lambda-equals-PSC} also gives rise
to the question whether $k$ for a given $PSC(\lambda)/PCE(\lambda)$
sentence $\varphi$, is a computable function of some (computable)
parameter of $\varphi$. We answer this in the negative by showing that
for every recursive function $\nu: \mathbb{N} \rightarrow \mathbb{N}$,
there is a $\Pi^0_2$ (resp. $\Sigma^0_2$) sentence that is
$PSC(\aleph_0)$ (resp. $PCE(\aleph_0)$) but that is not $PSC(k)$
(resp. $PCE(k)$) for any $k \leq \nu(|\varphi|)$, where $|\varphi|$
denotes the length of $\varphi$. Our proof uses an unpublished result
of Rossman~\cite{rossman-los-tarski} that gives a non-recursive lower
bound on the length of $\Pi^0_1$ sentences equivalent to sentences
defining hereditary classes (this strengthens a previous
non-elementary lower bound in the same
context~\cite{dawar-model-theory-large}). Our result thus shows that
if a sentence $\varphi$ is $PSC(k)/PCE(k)$, then the smallest such $k$
can be non-recursively larger than $|\varphi|$.

%% file: glt-k-for-theories-3.0.tex
\section{$\glt{k}$ for Theories}\label{section:glt-k-for-theories}

The case of theories (sets of sentences) for $\glt{k}$ turns out to be
much different than the case of sentences, both in terms of the
results in general and the methods to prove them.

\begin{theorem}[``Extensional'' results; Thm. 5.1.1, Prop. 5.1.4, Chp. 5]\label{theorem:ext-chars}
Let $k \in \mathbb{N}$ and $\lambda \ge \aleph_0$.
\begin{enumerate}[nosep]
\item A theory $T$ is $PCE(k)$ if, and only if, $T$ is equivalent to a
  theory of $\Pi^0_2$ sentences, all of which have $k$ universal
  quantifiers.\label{theorem:char-of-PCE(k)-theories}
\item A theory $T$ is $PCE(\lambda)$ if, and only if, $T$ is
  equivalent to a theory of $\Pi^0_2$
  sentences.\label{theorem:char-of-PCE(lambda)-theories}
\item The universal theory $T$ defining the class of undirected
  acyclic graphs is such that $T$ is $PCE(\aleph_0)$ but not $PCE(l)$
  for any $l \in \mathbb{N}$.
\end{enumerate}
\end{theorem}

\begin{theorem}[``Substructural'' results; Thm. 5.2.1, Chp. 5]\label{theorem:subst-char-for-theories}
Let $k \in \mathbb{N}$ and $\lambda \ge \aleph_1$.
\begin{enumerate}[nosep]
\item A theory $T$ is $PSC(\lambda)$ if, and only if, $T$ is
  equivalent to a theory of $\Sigma^0_2$
  sentences.\label{theorem:PSC(lambda)-char-for-theories}
\item If a theory $T$ is $PSC(\aleph_0)$, then $T$ is equivalent to a
  theory of $\Sigma^0_2$ sentences. The same therefore holds if $T$ is
  $PSC(k)$.\label{theorem:PSC-and-PSC(aleph_0)-for-theories}
\item The theory $T$ defining the class of graphs not containing for
  any $n$, a vertex cycle cover comprising cycles of length $\leq n$,
  is such that (i) $T$ is theory of $\Sigma^0_2$ sentences each of
  which has one existential quantifier, and (ii) $T$ is not
  $PSC(\aleph_0)$, and thus not $PSC(l)$ for any $l \in \mathbb{N}$.
\end{enumerate}
\end{theorem}

In contrast to Corollary~\ref{corollary:PSC-lambda-equals-PSC}, we see
that the ``infinite to finite collapse'' does not happen in the case
of theories. However a collapse does happen in the infinite realm: the
$PSC$ heirarchy collapses to the level $\aleph_1$, while the $PCE$
hierarchy collapses to the level $\aleph_0$. We do not know yet
however, whether $PSC(\aleph_0)$ collapses to the union of $PSC(k)$
over all $k$.

The proofs of the extensional characterizations above are similar to
the proof of the extensional part of Theorem~\ref{theorem:glt(k)}. The
substructural results require an altogether different approach, and we
establish them by using a characterization of $\Sigma^0_2$ theories in
terms of \emph{1-sandwiches}, the notion and the result both due to
Keisler~\cite{chang-keisler}. Given structures $\mf{A}$ and $\mf{B}$,
we say $\mf{B}$ is 1-sandwiched by $\mf{A}$ if there exist elementary
extensions $\mf{A}'$ and $\mf{B}'$ resp. of $\mf{A}$ and $\mf{B}$,
such that $\mf{A} \subseteq \mf{B}' \subseteq \mf{A}'$. A theory $T$
is \emph{preserved under 1-sandwiches} if for every model $\mf{A}$ of
$T$, if $\mf{B}$ is 1-sandwiched by $\mf{A}$, then $\mf{B}$ models
$T$. We show Theorem~\ref{theorem:subst-char-for-theories} by showing
that for any $\lambda$ and $k$, a $PSC(\lambda)/PSC(k)$ theory is
always preserved under 1-sandwiches. The idea, just as in the sketched
proof of Theorem~\ref{theorem:glt(k)}(\ref{theorem:glt(k)-ext}), is to
first prove the result for $\mu$-saturated structures, and then
``transfer it out'' to all structures.

\begin{lemma}[Lem. 5.2.5, Lem. 5.2.6, Chp. 5]\label{lemma:sandwich-lemma}
  The following are true.
  \begin{enumerate}[nosep]
  \item If $\mf{B}_1$ is 1-sandwiched by $\mf{A}_1$, and $\mf{A}$ is a
    $\mu$-saturated elementary extension of $\mf{A}_1$ for $\mu \ge
    \omega$, then there is an isomorphic copy $\mf{B}$ of $\mf{B}_1$
    such that $\mf{B}$ is sandwiched by $\mf{A}$.
  \item If $\mf{B}$ is 1-sandwiched by a $\mu$-saturated model
    $\mf{A}$ of a $PSC(\lambda)$ theory $T$, then $\mf{B}$ is a model
    of $T$.
  \end{enumerate}
\end{lemma}
    
Enroute proving Lemma~\ref{lemma:sandwich-lemma}, we show a crucial
result that we call the ``crux-determination'' lemma; we describe this
for $PSC(k)$ theories. Firstly we extend the notion of a $k$-crux from
sets to $k$-tuples in the natural way. The crux-determination lemma
characterizes the conditions under which the FO-type of a $k$-tuple
$\bar{a}$ of a structure $\mf{A}$ \emph{determines} a $k$-crux in a
model of a $PSC(k)$ theory $T$. That is, if a $k$-tuple $\bar{b}$ of a
structure $\mf{B}$ has the same FO-type as $\bar{a}$ in $\mf{A}$, then
$\mf{B}$ is a model of $T$ and $\bar{b}$ is a $k$-crux of $\mf{B}$
(w.r.t. $T$).  Towards the lemma, we introduce a key notion. For a
model $\mf{A}$ of a $PSC(k)$ theory $T$ and a $k$-tuple $\bar{a}$ of
$\mf{A}$, we say $\bar{a}$ is a \emph{distinguished $k$-crux} of
$\mf{A}$ if there is a $\mu$-saturated elementary extension $\mf{B}$
of $\mf{A}$ (whereby $\mf{B}$ models $T$) for some $\mu \ge \omega$,
such that $\bar{a}$ is a $k$-crux of $\mf{B}$ (and hence a $k$-crux of
$\mf{A}$).  We now have the following characterization.

\begin{lemma}[Crux-determination; Lem. 5.2.17, Chp. 5]\label{lemma:crux-determination}
Let $T$ be a theory that is $PSC(k)$. Then the universal type (and
hence the FO-type) of a $k$-tuple $\bar{a}$ of a structure $\mf{A}$
determines a $k$-crux in a model of $T$ if, and only if, $\mf{A}$ is a
model of $T$ and $\bar{a}$ is a distinguished $k$-crux of $\mf{A}$.
\end{lemma}

The notion of a distinguished $k$-crux and its characterization by
Lemma~\ref{lemma:crux-determination} above, turn out to play an
important role in obtaining a (conditional) refinement of
Theorem~\ref{theorem:subst-char-for-theories}(\ref{theorem:PSC-and-PSC(aleph_0)-for-theories}). The
latter, while showing that a $PSC(k)$ theory is always equivalent to a
$\Sigma^0_2$ theory, does not tell us anything about the number of
existential quantifiers appearing in the $\Sigma^0_2$ sentences of the
latter theory.  Given that a $PSC(k)$ sentence is always equivalent to
an $\exists^k \forall^*$ sentence, it is natural to ask if the same
holds for theories too. We answer this in the affirmative, conditioned
on a well-motivated hypothesis about $PSC(k)$ theories and
distinguished $k$-cruxes, that we state below. Observe that $PSC(k)$
sentences (singleton theories) for instance, satisfy this hypothesis.

\begin{hypothesis}\label{hypothesis:H}
Every model of a $PSC(k)$ theory contains a distinguished $k$-crux.
\end{hypothesis}

\begin{theorem}[Thm. 5.2.3, Chp. 5]\label{theorem:conditional-refinement}
Assume Hypothesis~\ref{hypothesis:H} holds. If a theory $T$ is
$PSC(k)$, then $T$ is equivalent to a theory of $\Sigma^0_2$
sentences, all of which have $k$ existential quantifiers.
\end{theorem}

The proof is amongst the most technically involved of the thesis, and
introduces a novel technique of getting a syntactically defined FO
theory equivalent to a given FO theory satisfying a semantic property,
\emph{by going outside of FO} (first presented
in~\cite{abhisekh-apal}).  The proof is in two parts:
\begin{enumerate}[nosep,leftmargin=*]
\item ``Going up'': We give a characterization of $PSC(k)$ theories in
  terms of sentences of the infinitary logic $\mathsf{L} = [\bigvee]
  \left[\exists^k \bigwedge\right] \Pi^0_1$, that consists of
  infinitary disjunctions of sentences obtained by taking the
  existential closure of infinitary conjunctions of $\Pi^0_1$
  formulae, all of whose free variables are among a given set of $k$
  variables.
\item ``Coming down'': We show that a sentence $\Phi$ of $\mathsf{L}$
  defines an elementary (i.e. FO definable) class if, and only if,
  $\Phi$ is equivalent to a countable subset of the set $\mc{A}(\Phi)$
  of suitably defined \emph{finite approximations} of $\Phi$. Each of
  these finite approximations would turn out to be an $\exists^k
  \forall^*$ FO sentence, proving
  Theorem~\ref{theorem:conditional-refinement}.
\end{enumerate}

We explain briefly the ideas involved in proving these parts.  For the
``Going up'' part, the non-trivial direction is showing that a
$PSC(k)$ theory $T$ is equivalent to an
$\mathsf{L}$-sentence. Consider the $\mathsf{L}$-sentence $\Phi$
obtained by taking the disjunction over all models $\mf{A}$ of $T$ and
all distinguished $k$-cruxes $\bar{a}$ of $\mf{A}$ (which exist by
Hypothesis~\ref{hypothesis:H}), of the existential closure of the
universal type of $\bar{a}$ in $\mf{A}$. That $T$ is equivalent to
$\Phi$ now follows from Lemma~\ref{lemma:crux-determination}.  For the
``Coming down'' part, consider the logic $\mathsf{L}_1 =
\left[\exists^k \bigwedge\right] {\fo}$ defined just as
$\left[\exists^k \bigwedge\right] \Pi^0_1$ above, by considering all
of FO instead of just $\Pi^0_1$. For $\Psi \in \mathsf{L}_1$, define
the set $\mc{A}(\Psi)$ of finite approximations of $\Psi$ as follows:
if $\Psi = \exists^k \bar{x} \bigwedge_{j \in J} \psi_{j}(\bar{x})$,
then $\mc{A}(\Psi) = \{\exists^k \bar{x} \bigwedge_{j \in J_1}
\psi_{j}(\bar{x}) \mid J_1 \subseteq_f J\}$ where $\subseteq_f$
denotes ``finite subset of''. And now for $\Phi \in \mathsf{L}$, if
$\Phi = \bigvee_{i \in I} \Psi_i$ where $\Psi_i \in \left[\exists^k
  \bigwedge\right] \Pi^0_1 \subseteq \mathsf{L}_1$, then define
$\mc{A}(\Phi) = \{ \bigvee_{i \in I_1} \psi_i \mid I_1 \subseteq_f I,
\psi_i \in \mc{A}(\Psi_i)\}$. Observe that each sentence in
$\mc{A}(\Phi)$ is equivalent to an $\exists^k \forall^*$ sentence.
The heart of the proof of the ``Coming down'' part is now
equivalence~(\ref{eq:1}) below for $\Phi = \bigvee_{i \in I} \Psi_i$
as above. Once~(\ref{eq:1}) is shown, the proof is completed by
rewriting the RHS of~(\ref{eq:1}) as a conjunction of disjuncts, and
then using (FO) Compactness to reduce each (infinite) disjunct down to
a finite subset of it.
\begin{eqnarray}
\Phi & \leftrightarrow & \bigvee_{i \in I} \bigwedge_{\psi \in
  \mc{A}(\Psi_i)} \psi \label{eq:1}
\end{eqnarray}
We conclude Part I by presenting our key result that enables us to
prove~(\ref{eq:1}) -- a Compactness theorem for
$\mathsf{L}_1$. The standard FO Compactness is a special
case of this theorem.
\begin{theorem}[Compactness for $\mathsf{L}_1$; Lem. 5.2.18, Lem. 5.2.21, Chp. 5]\label{theorem:compactness}
  Let $\Psi \in \mathsf{L}_1$ be given.
  \begin{enumerate}[nosep]
    \item Any model of $\Psi$ is also a model of $\mc{A}(\Psi)$.
    \item If every sentence of $\mc{A}(\Psi)$ is satisfiable, then
      $\Psi$ is satisfiable.
  \end{enumerate}
\end{theorem}

%% file: results-in-FMT.tex
\newcommand{\lebgenp}[3]{\ensuremath{\mc{L}\text{-}\mathsf{EBSP}(#1, #2, #3)}}
\vspace{1pt}
\begin{center}
\LARGE{\textsc{Part II: Finite Model Theory}}\label{section:results-in-FMT}
\end{center}
\vspace{10pt}

\input{need-for-new-classes-2.0}

\input{abstract-framework-3.0}

\input{applications-3.0}

\input{wqo}

\input{results-obtained-post-thesis-3.0}

%% file: need-for-new-classes-2.0.tex
We now turn our attention to finite structures. As the following
results show, $\glt{k}$ fails over the class of all finite structures,
and also over the classes of sparse graphs shown
in~\cite{dawar-pres-under-ext} to satisfy the {\lt} theorem. Below
$\cl{S}$-equivalent means ``equivalent over $\cl{S}$''.

\begin{prop}[Prop. 8.1.1, Chp. 8]\label{prop:failure-of-glt(k)-in-the-finite}
Let $\cl{S}$ be the class of all finite structures over a vocabulary
consisting of two binary predicates, one unary predicate and two
constants.  For each $k \ge 0$, there is an FO sentence $\varphi_k$
that is hereditary over $\cl{S}$ (hence $PSC(k)$ over $\cl{S}$), but
that is not $\cl{S}$-equivalent to any $\exists^k \forall^*$ sentence.
\end{prop}

\begin{prop}[Thm. 8.2.2, Chp. 8]\label{theorem:glt(k)-implies-bounded-ind-path-lengths}
  Let $\cl{S}$ be a hereditary graph class having unbounded induced
  path lengths. Then for each $k \ge 2$, there is an FO sentence
  $\varphi_k$ that is $PSC(k)$ over $\cl{S}$, but that is not
  $\cl{S}$-equivalent to any $\exists^k \forall^*$ sentence.
\end{prop}

For both results above, the proof idea is akin to the
Ehrenfeucht-Fr\"aiss\'e method for showing inexpressibility results in
FO: the sentence $\varphi_k$ is such that for each $n$, it has a model
and a non-model such that every $\exists^k \forall^n$ sentence true in
the model is also true in the non-model; then $\varphi_k$ cannot be
equivalent to any $\exists^k \forall^*$ sentence. Interestingly,
$\varphi_k$ itself turns out to be an $\exists^l \forall^*$ sentence
for $l > k$.

Proposition~\ref{theorem:glt(k)-implies-bounded-ind-path-lengths}
naturally leads us to consider \emph{dense structures}. We investigate
various classes of posets and dense graphs, all of active ongoing
interest. As mentioned in Section~\ref{section:overview}, our
investigations are done within an abstract framework that incorporates
the favourable structural and logical properties of the mentioned
classes. Our presentation of the framework
follows~\cite{abhisekh-csl17} which simplifies the more technical
presentation of the framework in Chapter 10.1.

%% file: abstract-framework-3.0.tex
\section{An Abstract Framework}\label{section:abstract-framework}

\noindent \tbf{A. $\mc{L}$-good tree representations:} We consider
classes of structures that admit tree representations in which the
leaf nodes represent simple structures and the internal nodes
represent operations on structures. Formally, our tree representations
are ordered trees over the finite alphabet $\sigmaint \cup \sigmaleaf$
where $\sigmaleaf$ denotes the structures labeling the leaves and
$\sigmaint$ denotes the operations. The operations can have fixed
arity or unbounded arity. The latter case is used to represent an
arbitrary number of iterations of a fixed arity operation; for
instance, the binary disjoint union operator has a natural extension
to an arbitrary arity version of it that iterates the binary disjoint
union over the inputs. We formalize these ideas by equipping our trees
with a ranking function $\rho: \sigmaint \rightarrow \mathbb{N}$ and a
subset $\sigmarank$ of $\sigmaint$ such that an operator $\mathsf{O}
\in \sigmarank$ has arity $\rho(\mathsf{O})$, and so does the fixed
arity operator corresponding to an operator $\mathsf{O} \in \sigmaint
\setminus \sigmarank$; then the allowed arities for $\mathsf{O} \in
\sigmaint \setminus \sigmarank$ belong to the set $\{ \rho(\mathsf{O})
+ i \cdot (\rho(\mathsf{O}) - 1) \mid i \in \mathbb{N}\}$.

Let $\mc{L}$ be one of the logics FO or MSO over a vocabulary $\tau$
and $\mc{L}[m]$ denote the sentences of $\mc{L}$ of quantifier rank
$m$. Let $\lequiv{m}$ denote the equivalence relation on
$\tau$-structures such that for $\tau$-structures $\mf{A}$ and
$\mf{B}$, we have $\mf{A} \lequiv{m} \mf{B}$ if and only if $\mf{A}$
and $\mf{B}$ agree on all sentences of $\mc{L}[m]$. Let $\Delta_{m,
  \mc{L}}$ be the set of equivalence classes of the $\lequiv{m}$
relation. For the purposes of our results, we consider
\emph{$\mc{L}$-good} operations that satisfy the properties stated
below:\\

\begin{enumerate}[nosep, leftmargin=*]
\item\label{monotonicity} Monotonicity: Let $\mathsf{O} \in \sigmaint$
  and let $n$ be the/an allowed arity of $\mathsf{O}$. Let
  $\mathsf{O}(\mf{A}_1, \ldots, \mf{A}_n)$ denote the output of
  $\mathsf{O}$ when fed with $\mf{A}_1, \ldots, \mf{A}_n$ as
  inputs. Let $\hookrightarrow$ denote ``isomorphically embeddable''.
  \begin{enumerate}
    \item $\mf{A}_i \hookrightarrow \mathsf{O}(\mf{A}_1, \ldots,
      \mf{A}_n)$ for $i \in \{1, \ldots, n\}$.
    \item If $\mf{B}_i \hookrightarrow \mf{A}_i$ for $i \in \{1,
      \ldots, n\}$, then $\mathsf{O}(\mf{B}_1, \ldots, \mf{B}_n)
      \hookrightarrow \mathsf{O}(\mf{A}_1, \ldots, \mf{A}_n)$.
    \item\label{embedding} Suppose $\mathsf{O} \in \sigmaint \setminus
      \sigmarank$, $r = \rho(\mathsf{O}), n = r + q \cdot (r - 1)$ and
      $i = r + j \cdot (r - 1)$ for some $j \in \{0, \ldots,
      q-1\}$. Then $\mathsf{O}(\mf{A}_1, \ldots, \mf{A}_i, \mf{A}_{i +
        r}, \ldots, \mf{A}_n) \hookrightarrow \mathsf{O}(\mf{A}_1,
      \ldots, \mf{A}_n)$.
  \end{enumerate}
  
\item Feferman-Vaught composition (FVC): The $\mc{L}$-FVC property of
  an operation $\mathsf{O} \in \sigmaint$ intuitively states that the
  $\lequiv{m}$-equivalence classes of the inputs to $\mathsf{O}$
  \emph{determine} the $\lequiv{m}$-equivalence class of its
  output. Formally, there is a \emph{composition function} $f_{m,
    \mathsf{O}} : (\Delta_{m, \mc{L}})^{\rho(\mathsf{O})} \rightarrow
  \Delta_{m, \mc{L}}$ such that the following hold. Let
  $\delta(\mf{A})$ denote the $\lequiv{m}$-equivalence class of
  $\mf{A}$.
  \begin{itemize}
    \item If $\mathsf{O} \in \sigmarank$, then
      $\delta(\mathsf{O}(\mf{A}_1, \ldots, \mf{A}_n)) = f_{m,
      \mathsf{O}}(\delta(\mf{A}_1), \ldots, \delta(\mf{A}_n)))$
      where $n = \rho(\mathsf{O})$.
    \item If $\mathsf{O} \in \sigmaint \setminus \sigmarank$ and $r,
      n, i, j$ are as in point (\ref{embedding}) above, then
      $\delta_m(\mathsf{O}(\mf{A}_1, \ldots, \mf{A}_n)) = \chi_q$
      where $\chi_0 = f_{m, \mathsf{O}}(\delta(\mf{A}_1), \ldots,
      \delta(\mf{A}_r))$ and $\chi_{j+1} = f_{m, \mathsf{O}}(\chi_j,
      \delta(\mf{A}_{i+1}), \ldots, \delta(\mf{A}_{i + r - 1}))$.\\
  \end{itemize}
    
\end{enumerate}

The above properties are satisfied by a variety of operations as we
will see later. As quick examples, disjoint union satisfies the above
properties for MSO, as does Cartesian product for FO. For a set
$\sigmaint$ of $\mc{L}$-good operations and a tree $\tree{t}$ over
$\sigmaint \cup \sigmaleaf$, let $\mf{A} = \str{\tree{t}}$ be the
natural structure associated with $\tree{t}$, obtained by a
``bottom-up evaluation'' in the latter. We then say $\tree{t}$ is an
$\mc{L}$-good \emph{tree representation} of $\mf{A}$ over $\sigmaint
\cup \sigmaleaf$ and that $\mathsf{Str}$ is an $\mc{L}$-good
\emph{representation map}. We say a class $\cl{S}$ of structures
\emph{admits an $\mc{L}$-good tree representation} if there exist
$\sigmaint$ and $\sigmaleaf$ such that for each $\mf{A} \in \cl{S}$,
there is an $\mc{L}$-good tree representation of $\mf{A}$ over
$\sigmaint \cup \sigmaleaf$. The following theorem is at the heart of
most of the results in the remainder of this report. It shows shows
why $\mc{L}$-good tree representations are called so. The theorem is a
joint presentation of Theorems 10.1.1 and 10.4.11 from Chapter
10. (The latter theorems together are actually slighty more general.)

\begin{theorem}\label{theorem:abstract-tree-theorem}
Let $\cl{S}$ be a class of structures that admits an $\mc{L}$-good
tree representation. Let $\mc{T}$ be a class of $\mc{L}$-good tree
representations of the structures of $\cl{S}$ over some alphabet
$\sigmaleaf \cup \sigmaint$ and let $\mathsf{Str}: \mc{T} \rightarrow
\cl{S}$ be the associated $\mc{L}$-good representation map. Suppose
$\mc{T}$ is a regular language of trees. Then there exist computable
functions $\eta_1, \eta_2: \mathbb{N} \rightarrow \mathbb{N}$ such
that for each $\tree{t} \in \mc{T}$ and $m \in \mathbb{N}$, we have
the following:
\begin{enumerate}[nosep]
\item \emph{(Height reduction)} There exists a subtree $\tree{s}_1$ of
  $\tree{t}$, of height $\leq \eta_1(m)$, such that (i) $\tree{s}_1
  \in \mc{T}$, (ii) $\str{\tree{s}_1} \hookrightarrow \str{\tree{t}}$,
  and (iii) $\str{\tree{s}_1} \lequiv{m} \str{\tree{t}}$.
  \label{lemma:abstract-tree-lemma-height-reduction}
\item \emph{(Degree reduction)} There exists a subtree $\tree{s}_2$ of
  $\tree{t}$, of degree $\leq \eta_2(m)$, such that (i) $\tree{s}_2
  \in \mc{T}$, (ii) $\str{\tree{s}_2} \hookrightarrow \str{\tree{t}}$,
  and (iii) $\str{\tree{s}_2} \lequiv{m} \str{\tree{t}}$.
  \label{lemma:abstract-tree-lemma-degree-reduction}
\end{enumerate}
\end{theorem}

Indeed Theorem~\ref{theorem:abstract-tree-theorem} shows that for
$\cl{S}$ as in the theorem, for any structure $\mf{A}$ in $\cl{S}$, a
degree and height reduction of a tree representation $\tree{t}$ of
$\mf{A}$ given by $\mathsf{Str}$ yields a computably small subtree of
$\tree{t}$ that represents a small $\mc{L}[m]$-similar substructure of
$\mf{A}$ in $\cl{S}$. We use this crucially in our model-theoretic
results. Theorem~\ref{theorem:abstract-tree-theorem} turns out to also
have important algorithmic and ``conceptual'' consequences
(Section~\ref{section:post-thesis}) due to the constructive nature of
its proof that we sketch now.  We make an important use of a
composition lemma for ordered trees given by
Lemma~\ref{lemma:mso-composition-lemma-for-ordered-trees}. Let
$\tree{t} \in \mc{T}$ be given. Since $\mc{T}$ is regular, it is
definable in MSO~\cite{tata} by a sentence of rank say $n$.
 
\begin{enumerate}[nosep, leftmargin=*]
\item Height reduction: Suppose there is a long root-to-leaf path in
  $\tree{t}$. We label each node $a$ of this path with the pair
  $(\delta_1, \delta_2)$ where $\delta_1$ is the $\lequiv{m}$ class of
  the structure $\mf{A}_a$ represented by $\tree{t}_{\ge a}$ which is
  the subtree of $\tree{t}$ rooted at $a$ ($\tree{t}_{\ge a}$ need not
  be in $\mc{T}$), and $\delta_2$ is the $\mequiv{n}$ class of
  $\tree{t}_{\ge a}$ itself. Given that the indices of the
  $\lequiv{m}$ and $\mequiv{n}$ relations over all finite structures
  are finite (and bounded by computable functions of $m$ and $n$
  resp.~\cite{libkin}), the number of pairs $(\delta_1, \delta_2)$ is
  finite (and bounded by a computable function of $m$ and $n$). Then
  some such pair repeats at a node $a$ and a descendent $b$ of it
  along the path. We then replace $\tree{t}_{\ge a}$ with
  $\tree{t}_{\ge b}$ to get a proper subtree $\tree{t}_1$ of
  $\tree{t}$.  Since $\tree{t}_{\ge b} \mequiv{n} \tree{t}_{\ge a}$ it
  follows by Lemma~\ref{lemma:mso-composition-lemma-for-ordered-trees}
  that $\tree{t}_1 \mequiv{n} \tree{t}$; then $\tree{t}_1 \in \mc{T}$
  since $\tree{t} \in \mc{T}$. Again since $\mf{A}_b \lequiv{m}
  \mf{A}_a$ and since the operations of $\sigmaint$ are $\mc{L}$-good,
  we have $\str{\tree{t}_1} \hookrightarrow \str{\tree{t}}$ and
  $\str{\tree{t}_1} \lequiv{m} \str{\tree{t}}$. Iterating, we
  eventually get the desired subtree $\tree{s}_1$.
\item Degree reduction: We illustrate our reasoning for the case when
  for each $\mathsf{O} \in \sigmaint \setminus \sigmarank$, we have
  $\rho(\mathsf{O}) = 2$.  Let $a$ be a node of $\tree{t}$ of large
  degree, say $r$; then the operation labeling it is in $\sigmaint
  \setminus \sigmarank$. Let $\tree{z} = \tree{t}_{\ge a}$. For $i \in
  \{1, \ldots, r-1\}$, let $\tree{x}_i$, resp. $\tree{y}_i$, be the
  subtree of $\tree{z}$ obtained by retaining the first $i$, resp. the
  last $r - i$, child subtrees of the root of $\tree{z}$ and deleting
  the rest; then $\tree{z}$ is the tree $\tree{x}_i \odot \tree{y}_i$
  obtained by merging $\tree{x}_i$ and $\tree{y}_i$ at their roots
  (and in that order). Label each $\tree{x}_i$ with the pair
  $(\delta_1, \delta_2)$ as described above; so $\delta_1$ is the
  $\lequiv{m}$ class of the structure represented by $\tree{x}_i$ and
  $\delta_2$ the $\mequiv{n}$ class of $\tree{x}_i$ itself. Since $r$
  is large, such a pair repeats for some $l, k \in \{1, \ldots, r-1\},
  l < k$. Then consider the tree $\tree{z}_2 = \tree{x}_l \odot
  \tree{y}_k$ and let $\tree{t}_2$ be the proper subtree of $\tree{t}$
  obtained by replacing $\tree{z}$ with $\tree{z}_2$ in $\tree{t}$.
  By Lemma~\ref{lemma:mso-composition-lemma-for-ordered-trees},
  $\tree{z}_2 \mequiv{n} \tree{z}$, whereby $\tree{t}_2 \mequiv{n}
  \tree{t}$; then $\tree{t}_2 \in \mc{T}$. Since the operations of
  $\sigmaint$ are $\mc{L}$-good, we have $\str{\tree{z}_2}
  \hookrightarrow \str{\tree{z}}$ and $\str{\tree{z}_2} \lequiv{m}
  \str{\tree{z}}$; then $\str{\tree{t}_2} \hookrightarrow
  \str{\tree{t}}$ and $\str{\tree{t}_2} \lequiv{m}
  \str{\tree{t}}$. Iterating, we eventually get the desired subtree
  $\tree{s}_2$.
\end{enumerate}

\input{ebsp-3.0}

%% file: ebsp-3.0.tex
\vspace{5pt}
\noindent \tbf{B. The $\mc{L}$-Equivalent Bounded Substructure Property}:
The following abstract property of finite structures formalizes the
implication of Theorem~\ref{theorem:abstract-tree-theorem}, discussed
above. (This property was first introduced in~\cite{abhisekh-mfcs} for
FO.)

\begin{defn}[$\lebsp{\cl{S}, k}$; Def. 9.1, Chp. 9]\label{defn:lebsp}
Let $\cl{S}$ be a class of structures and $k \in \mathbb{N}$.  We say
that $\cl{S}$ satisfies the \emph{$\mc{L}$-Equivalent Bounded
Substructure Property} for parameter $k$, abbreviated
\emph{$\lebsp{\cl{S}, k}$}, if there exists a function
$\vartheta: \mathbb{N} \rightarrow \mathbb{N}$ such that for each
$m \in \mathbb{N}$, for each structure $\mf{A}$ of $\cl{S}$ and for
each subset $W$ of at most $k$ elements from $\mf{A}$, there exists a
structure $\mf{B}$ such that (i) $\mf{B} \in
\cl{S}$, (ii) $\mf{B} \subseteq \mf{A}$, (iii) the elements of $W$ are
contained in $\mf{B}$, (iv) $|\mf{B}| \leq \vartheta(m)$, and (v)
$\mc{B} \lequiv{m} \mf{A}$. We call $\vartheta$ a \emph{witness
function} for $\lebsp{\cl{S}, k}$.
\end{defn}

The above definition does not insist on the computability of the
witness function; we present scenarios later
(Section~\ref{section:wqo}) where the witness functions are
necessarily uncomputable. However, for the classes we consider in this
section, we have the following result that
Theorem~\ref{theorem:abstract-tree-theorem} entails.

\begin{prop}[Lem. 10.1.2, Chp. 10]\label{prop:L-good-tree-reps-and-ebsp}
For $k \in \mathbb{N}$ and a class $\cl{S}$ of structures, let
$\cl{S}_k$ be the class of structures obtained by labeling (possibly
partially) the elements of the structures of $\cl{S}$ with labels from
$\{1, \ldots, k\}$. Suppose $\cl{S}_k$ admits an $\mc{L}$-good tree
representation with a computable $\mc{L}$-good representation
map. Then $\lebsp{\cl{S}, k}$ holds with a computable witness
function.
\end{prop}

Section~\ref{section:applications} discusses a number of concrete
instances where the premises of
Proposition~\ref{prop:L-good-tree-reps-and-ebsp} are satisfied, wherby
these instances satisfy $\lebsp{\cdot, k}$. As mentioned in
Section~\ref{section:overview}, $\lebsp{\cdot, k}$ can be seen as a
finitary analogue of the $\refdls$ property; then a class satisfying
$\lebsp{\cdot, k}$ satisfies (a finitary adaptation of) the $\refdls$
theorem. We remark that there has been no study of the $\refdls$ (or
adaptations of it) over finite structures, except for~\cite{grohe-dls}
which proves a number of negative results concerning this theorem over
the class of all finite structures.

\vspace{5pt}
\noindent \tbf{C. $\lebsp{\cl{S}, k}$ entails $\glt{k}$ and $\hpt$}: We say an $\mc{L}$ sentence $\varphi$ is $PSC(k)$ over a class
$\cl{S}$ if the class of models of $\varphi$ in $\cl{S}$ is $PSC(k)$
relativized to $\cl{S}$. We say $\mc{L}$-$\glt{k}$ holds over $\cl{S}$
if for all $\mc{L}$ sentences $\varphi$, we have $\varphi$ is $PSC(k)$
over $\cl{S}$ if, and only if, $\varphi$ is $\cl{S}$-equivalent to an
$\exists^k\forall^*$ FO sentence.

\begin{theorem}[Thm. 9.1.2, Chp. 9]\label{theorem:lebsp-implies-glt(k)}
Let $\cl{S}$ be a class of finite structures and $k \in \mathbb{N}$ be
such that $\lebsp{\cl{S}, k}$ holds. Then $\mc{L}$-$\glt{k}$, and
hence $\glt{k}$ and the {\lt} theorem, hold over $\cl{S}$.  Further,
if there is a computable witness function for $\lebsp{\cl{S}, k}$,
then the translation from an $\mc{L}$ sentence that is $PSC(k)$ over
$\cl{S}$, to an $\cl{S}$-equivalent $\exists^k\forall^*$ sentence, is
effective.
\end{theorem}

The key idea of the proof is to construct for a given $PSC(k)$
sentence $\varphi$, an $\exists^k \forall^*$ sentence that checks in
any given structure $\mf{A}$, the existence of a set $W$ of $\leq k$
elements such that the truth of $\varphi$ in the substructures of
$\mf{A}$ in $\cl{S}$, that contain $W$, and that are of \emph{bounded
size}, itself suffices to ascertain the truth of $\varphi$ in
$\mf{A}$. Given that $\lebsp{\cl{S}, k}$ is true, if $\vartheta$ is a
witness function and $m$ is the rank of $\varphi$, then one sees that
the mentioned bound can indeed be taken to be $\vartheta(m)$.

Using similar ideas as above in ``dual'' form, we show that
$\lebsp{\cl{S}, k}$ entails a generalization of the $\hpt$. The $\hpt$
characterizes preservation under homomorphisms in terms of
existential-positive sentences which are FO sentences built up from
positive atomic formulae using conjunctions, disjunctions and
existential quantifications. Towards our result, we
define \emph{$k$-ary homomorphic covers} and \emph{preservation under
$k$-ary homomorphic coverings}, akin to the notions in
Definition~\ref{defn:k-ary-covered-ext}.

\begin{defn}[Defn. 11.3.2, Defn. 11.3.4, Chp. 11]\label{defn:k-ary-hom-cover}

Given a $\tau$-structure $\mf{A}$, a non-empty collection $\mc{R}$ of
$\tau_k$-structures (expansions of $\tau$-structures with $k$
constants) is called a \emph{$k$-ary homomorphic cover} of $\mf{A}$ if
for every $k$-tuple $\bar{a}$ of $\mf{A}$, there is a homomorphism
$h_{\bar{a}}: (\mf{B}, \bar{b}) \rightarrow (\mf{A}, \bar{a})$ for
some $(\mf{B}, \bar{b}) \in \mc{R}$. The set
$\{h_{\bar{a}} \mid \bar{a} \text{ is a }k\text{-tuple of }\mf{A}\}$
is called a \emph{$k$-ary homomorphic covering from $\mc{R}$ to
$\mf{A}$}.  For an $\mc{L}$ sentence $\varphi$, if $\mc{M}$ is the
class of expansions of the models of $\varphi$ with $k$ constants,
then we say $\varphi$ is \emph{preserved under $k$-ary homomorphic
coverings}, in short \emph{$\varphi$ is $\hpc{k}$}, if for every
collection $\mc{R}$ of structures of $\mc{M}$, if there is a $k$-ary
homomorphic covering from $\mc{R}$ to $\mf{A}$, then $\mf{A}$ models
$\varphi$.
\end{defn}

We say that a class $\cl{S}$ satisfies the \emph{generalized
$\hpt$ for $\mc{L}$ and $k$}, denoted $\lghpt{k}$, if the following is
true: an $\mc{L}$ sentence $\varphi$ is $\hpc{k}$ over $\cl{S}$ if,
and only if, $\varphi$ is $\cl{S}$-equivalent to a
\emph{$(\forall^k \exists^*)$-positive} (FO) sentence which is sentence
having the form $\forall x_1 \ldots \forall x_k \psi(x_1, \ldots,
x_k)$ where $\psi(x_1, \ldots, x_k)$ is an existential positive
formula. We now show $\lebsp{\cl{S}, k}$ entails $\lghpt{k}$ for all
$k$, by showing that a homomorphic version of $\lebsp{\cl{S}, k}$
``interpolates'' the said implication.

\begin{defn}[$\hlebsp{\cl{S}}{k}$; Def. 11.3.5, Chp. 11]
A class $\cl{S}$ satisfies the \emph{homomorphic
$\mc{L}$-$\mathsf{EBSP}$} for parameter $k$, abbreviated
\emph{$\hlebsp{\cl{S}}{k}$}, if there is a function
$\vartheta: \mathbb{N} \rightarrow \mathbb{N}$ such that for each
$m \in \mathbb{N}$, for each structure $\mf{A}$ of $\cl{S}$ and for
every $k$-tuple $\bar{a}$ from $\mf{A}$, there exists
$\mf{B} \in \cl{S}$ and a $k$-tuple $\bar{b}$ of $\mf{B}$ such that
(i) there is a homomorphism $h:$ $ (\mf{B}, \bar{b}) $ $\rightarrow
(\mf{A}, \bar{a})$, (ii) $|\mf{B}| \leq \vartheta(m)$, and (iii)
$\mf{B} \lequiv{m} \mf{A}$.  We call $\vartheta$ a \emph{witness
function} of $\hlebsp{\cl{S}}{k}$.
\end{defn}

\begin{theorem}[Thm. 11.3.7, Chp. 11]\label{theorem:h-ebsp-implies-ghpt(k)}
Let $\cl{S}$ be a class of finite structures and $k \in \mathbb{N}$ be
such that $\hlebsp{\cl{S}}{k}$ holds. Then $\lghpt{k}$, and hence
$\hpt$, hold over $\cl{S}$.  Further, if there is a computable witness
function for $\hlebsp{\cl{S}}{k}$, then the translation from an
$\mc{L}$ sentence that is $\hpc{k}$ over $\cl{S}$ to an
$\cl{S}$-equivalent $(\forall^k\exists^*)$-positive sentence, is
effective.  The above results also hold with $\lebsp{\cl{S}, k}$ in
place of $\hlebsp{\cl{S}}{k}$ (as the former entails the latter).
\end{theorem}

%% file: applications-3.0.tex
\section{Applications}\label{section:applications}

In this section, we show that various classes of dense structures,
specifically posets and recently defined subclasses of bounded
clique-width graphs, fall within the abstract framework described
above and are hence model-theoretically very well-behaved: they admit
the FVC property for the operations that construct their structures,
and effective versions of the $\refdls$, $\lglt{k}$ and $\lghpt{k}$
theorems for all $k$ and for $\mc{L}$ as FO and MSO.

\vspace{4pt}

\noindent \tbf{A. Words, trees (unordered, ordered, ranked or
  partially ranked) and nested words:} A tree (of any of the above
kinds) over a finite alphabet $\Sigma$ has a natural $\mc{L}$-good
tree representation; we describe this for an ordered partially ranked
$\Sigma$-tree whose ranking function is $\nu: X \rightarrow
\mathbb{N}$ where $X \subseteq \Sigma$. The tree representation has
the following parameters: $\sigmaleaf = \Sigma$, $\sigmaint =
\{\mathsf{O}_a \mid a \in \Sigma\}$, $\sigmarank = \{\mathsf{O}_a \mid
a \in X\}$, and $\rho: \sigmaint \rightarrow \mathbb{N}$ is such that
$\rho(\mathsf{O}_a) = \nu(a)$ if $a \in X$, else $\rho(\mathsf{O}_a) =
2$. Here $\mathsf{O}_a$ takes in a sequence of $n$ trees as input,
makes them the child subtrees in that order, of a new root node
labeled $a$ and outputs the resulting tree. The monotonicity
properties (Section~\ref{section:abstract-framework}.A) of
$\mathsf{O}_a$ are easy to see. The FVC property follows from
Lemma~\ref{lemma:mso-composition-lemma-for-ordered-trees}. We first
introduce some terminology.  For an alphabet $\Omega$, given ordered
$\Omega$-trees $\tree{t}, \tree{s}$ and a non-root node $a$ of
$\tree{t}$, the \emph{join of $\tree{s}$ to $\tree{t}$ to the right of
  $a$}, denoted $\tree{t} \cdot^{\rightarrow}_a \tree{s}$, is defined
(upto isomorphism) as the tree obtained by making $\tree{s}$ as a new
child subtree of the parent of $a$ in $\tree{t}$, at the successor
position of the position of $a$ among the siblings of $a$ in
$\tree{t}$.  Similarly define $\tree{t} \cdot^{\leftarrow}_a \tree{s}$
(joining to the left of $a$) and $\tree{t} \cdot^{\uparrow}_a
\tree{s}$ (joining below $a$).

\begin{lemma}[Composition lemma for ordered trees; Lem. 10.2.3, Chp. 10]\label{lemma:mso-composition-lemma-for-ordered-trees}
Given a finite alphabet $\Omega$, let ${\tree{t}}_i, \tree{s}_i$ be
non-empty ordered $\Omega$-trees, and let $a_i$ be a non-root node of
$\tree{t}_i$, for each $i \in \{1, 2\}$. Let $m \ge 2$ and suppose
that $({\tree{t}}_1, a_1) \lequiv{m} ({\tree{t}}_2, a_2)$ and
${\tree{s}}_1 \lequiv{m} {\tree{s}}_2$. Then $(({\tree{t}}_1
\cdot^{\rightarrow}_{a_1} {\tree{s}}_1), a_1) \lequiv{m}
(({\tree{t}}_2 \cdot^{\rightarrow}_{a_2} {\tree{s}}_2), a_2)$. The
result also holds if we replace $\cdot^{\rightarrow}$ with
$\cdot^{\leftarrow}$ or $\cdot^{\uparrow}$.
\end{lemma}

Nested words have natural representations using trees of our kind. The
non-trivial part here is showing the FVC property and this follows
from Lemma~\ref{lemma:nested-words-composition-lemma}. For given
nested words $\nesword{u}$ and $\nesword{v}$, let
$\nesword{u}\uparrow_{e}\nesword{v}$ denote the nested word obtained
by inserting $\nesword{v}$ in $\nesword{u}$ at a position $e$ of the
latter.

\begin{lemma}[Composition lemma for nested words; Lem. 10.2.6, Chp. 10]\label{lemma:nested-words-composition-lemma}
For a finite alphabet $\Sigma$, let $\nesword{u}_i, \nesword{v}_i$ be
nested $\Sigma$-words and let $e_i$ be a position in $\nesword{u}_i$
for $i \in \{1, 2\}$. Then given $m \in \mathbb{N}$, if
$(\nesword{u}_1, e_1) \lequiv{m} (\nesword{u}_2, e_2)$ and
$\nesword{v}_1 \lequiv{m} \nesword{v}_2$, then
$(\nesword{u}_1\uparrow_{e_1}\nesword{v}_1) \lequiv{m}
(\nesword{u}_2\uparrow_{e_2}\nesword{v}_2)$.
\end{lemma}

\noindent \tbf{B. $n$-partite cographs}: These recently defined
graphs~\cite{shrub-depth} are subclasses of bounded clique-width
graphs, that generalize a number of graph classes: threshold graphs,
cographs, graph classes of bounded tree-depth and those of bounded
shrub-depth.  An $n$-partite cograph $G = (V, E)$ is a graph that is
built up from the point graphs corresponding to the vertices of $V$,
labeled with labels from $\left[n\right] = \{1, \ldots, n\}$, using
operations $\mathsf{O}_f$ defined as follows for functions
$f:\left[n\right]^2 \rightarrow \{0, 1\}$: the operation
$\mathsf{O}_f$ takes in $p \ge 2$ graphs $G_1, \ldots, G_p$ that are
vertex-labeled with labels from $\left[n\right]$, and produces a graph
that is their disjoint union alongwith the addition of all edges
between vertices of $G_i$ with label $l$ and vertices of $G_j$ with
label $k$ where $1 \leq i < j \leq p, 1 \leq l, k \leq n$, and $f(l,
k) = 1$. One sees that $n$-partite cographs are exactly the subclass
of graphs of NLC-width $\leq n$ (and hence clique-width $\leq n$) that
are defined without relabelings. Further, these graphs fall within our
abstract framework with $\sigmaleaf = \left[n\right]$, $\sigmaint = \{
\mathsf{O}_f \mid f:\left[n\right]^2 \rightarrow \{0, 1\}\},
\sigmarank = \emptyset$ and $\rho$ as the constant 2.  The
monotonicity properties are easy to see; the FVC property follows by
Lemma~\ref{lemma:composition-lemma-for-n-partite-cographs} below and
the fact that $\mathsf{O}_f(G_1, \ldots, G_n) = \mathsf{O}_f(H_{n-2},
G_n)$ where $H_1 = \mathsf{O}_f(G_1, G_2)$ and $H_i =
\mathsf{O}_f(H_{i-1}, G_{i+1})$ for $1 \leq i \leq n-2$.

\begin{lemma}[Composition lemma for $\mathsf{O}_f$; Lem. 10.3.2, Chp. 10]\label{lemma:composition-lemma-for-n-partite-cographs}
  For $n \in \mathbb{N}$, let $G_i$ and $H_i$ be graphs whose vertices
  are labeled with labels from $\left[n\right]$, for $i \in \{1,
  2\}$. If $G_1 \lequiv{m} G_2$ and $H_1 \lequiv{m} H_2$, then
  $\mathsf{O}_f(G_1, H_1) \lequiv{m} \mathsf{O}_f(G_2, H_2)$ for each
  function $f: \left[n\right]^2 \rightarrow \{0, 1\}$ and $m \in
  \mathbb{N}$.
\end{lemma}

\noindent \tbf{C. Classes generated using set theoretic and logical
  operations}: We now present a number of methods of generating
classes that satisfy $\lebsp{\cdot, k}$ from those known to satisfy
the latter, thereby preserving the model-theoretic properties of the
latter entailed by $\lebsp{\cdot, k}$ (Chapter 10.4). For $i \in \{1,
2\}$, suppose $\lebsp{\cl{S}_i, k_i}$ is true with witness function
$\vartheta_i$. Then each of the following classes satisfy
$\lebsp{\cdot, k}$ for the $k$ and witness function $\vartheta$
mentioned.
\begin{enumerate}[nosep]
\item Any hereditary subclass of $\cl{S}_i$, with $k = k_i$ and
  $\vartheta = \vartheta_i$
\item The union $\cl{S}_1 \cup \cl{S}_2$, with $k = \text{min}(k_1,
  k_2)$ and $\vartheta = \text{max}(\vartheta_1, \vartheta_2)$
\item The intersection $\cl{S}_1 \cap \cl{S}_2$, with $k = k_2$ and
  $\vartheta = \vartheta_2$ if $\cl{S}_1$ is hereditary, and $k =
  \text{max}(k_1, k_2)$ and $\vartheta = \text{max}(\vartheta_1,
  \vartheta_2)$ if both $\cl{S}_1, \cl{S}_2$ are hereditary
\item Any $\mc{L}[r]$ definable subclass of $\cl{S}_i$, with $k = k_i$
  and $\vartheta(m) = \vartheta_i(r)$ if $m \leq r$ else $\vartheta(m)
  = \vartheta_i(m)$.
\end{enumerate}

We now look at classes generated using operations that are
``implementable'' using quantifier-free translation
schemes~\cite{makowsky}. Specifically, we consider such translation
schemes that ``act on'' the \emph{$n$-disjoint sum} of input
structures $\mf{A}_1, \ldots, \mf{A}_n$ or the \emph{$n$-copy} of an
input structure $\mf{A}$. The former is the structure obtained by
expanding the disjoint union of $\mf{A}_1, \ldots, \mf{A}_n$ with
fresh unary predicates $P_1, \ldots, P_n$ where $P_i$ is interpreted
as the universe of $\mf{A}_i$ for $i \in \{1, \ldots, n\}$. The latter
is the structure obtained by equipping the $n$-disjoint sum of $n$
isomorphic copies of $\mf{A}$ with a binary relation that relates
corresponding elements in these isomorphic copies. Let $\mathsf{O}$ be
an $n$-ary operation implemented by the above mentioned kinds of
translation schemes. Define the \emph{dimension} of $\mathsf{O}$ to be
the minimum dimension of its implementing translation schemes (the
dimension of the latter is the number of free variables in its
universe-defining formula).  Call $\mathsf{O}$ as ``sum-like'' if its
dimension is one, else call it ``product-like''. For example, disjoint
union and the operator $\mathsf{O}_f$ of
Lemma~\ref{lemma:composition-lemma-for-n-partite-cographs} are
sum-like, whereas Cartesian and tensor products are product-like. We
now have the following.

\begin{prop}[Cor. 10.4.7, Chp. 10]\label{prop:closure-of-lebsp-under-1-step-operations}
Let $\cl{S}_1, \ldots, \cl{S}_n$ and $\cl{S}$ be classes of structures
and let
$\mathsf{O}: \cl{S}_1 \times \cdots \times \cl{S}_n \rightarrow
\cl{S}$ be a surjective $n$-ary operation that is implementable using
a quantifier-free translation scheme of the kind mentioned above. Let
the dimension of $\mathsf{O}$ be $t$.
\begin{enumerate}[nosep]
\item If $\lebsp{\cl{S}_i, k_i}$ is true for $k_i \in \mathbb{N}$ for
  each $i \in \{1, \ldots, n\}$, then so is $\lebsp{\cl{S}, l}$, for
  $l = \text{min}\{k_i \mid i \in \{1, \ldots, n\}\}$, whenever
  $\mathsf{O}$ is sum-like.
\item If $\febsp{\cl{S}_i, k_i \cdot t}$ is true for $k_i \in
  \mathbb{N}$ for each $i \in \{1, \ldots, n\}$, then so is
  $\febsp{\cl{S}, l}$, for $l = \text{min}\{k_i \mid i \in \{1,
  \ldots, n\}\}$, whenever $\mathsf{O}$ is product-like.
\end{enumerate}
In the implications above, if there are computable witness functions
for each of the conjuncts in the antecedent, then there is a
computable witness function for the consequent as well.
\end{prop}

We now observe that
Proposition~\ref{prop:closure-of-lebsp-under-1-step-operations} in
conjunction with the set theoretic closure properties above, shows
that finite unions of classes obtained by applying finite compositions
of the operations of above kind, to a given class $\cl{S}$ of
structures, preserve the $\lebsp{\cdot, \cdot}$ property of
$\cl{S}$. However, given that taking (even binary) unions can in
general increase the value of the witness function, it is unclear if
infinite unions of the kind mentioned would preserve $\lebsp{\cdot,
  \cdot}$. We show that if the infinite unions are ``regular'', then
$\lebsp{\cdot, 0}$ indeed remains preserved. More precisely, consider
a set $\mathsf{Op}$ of $\mc{L}$-good operations and let $\mc{T}$ be a
class of trees over $\mathsf{Op}$ in which the leaf nodes are labeled
with a symbol $\diamond$ that acts a ``place holder'' for an input
structure. Each tree $\tree{t} \in \mc{T}$ can be seen as an operation
itself, with inputs fed at the leaves and output obtained at the
root. Given a class $\cl{S}$, let $\tree{t}(\cl{S})$ denote the class
obtained by ``applying'' $\tree{t}$ to the structures of $\cl{S}$. By
extension, let $\mc{T}(\cl{S}) = \bigcup_{\tree{t} \in \mc{T}}
\tree{t}(\cl{S})$. We now show the following result using similar
ideas as for Theorem~\ref{theorem:abstract-tree-theorem}.

\begin{theorem}\label{theorem:op-tree-languages}
Let $\mc{T}$ as described above be a regular language of trees. If
$\lebsp{\cl{S}, 0}$ is true (with a computable witness function), then
so is $\lebsp{\mc{T}(\cl{S}), 0}$.
\end{theorem}

Using the methods above, we get a wide array of classes satisfying
$\lebsp{\cdot, k}$. Classes 1-11 in Table~\ref{table:logical-fractals}
of Section~\ref{section:post-thesis} are examples (these in fact
satisfy a strengthened version of $\lebsp{\cdot, k}$).

%% file: wqo.tex
\section{Well-quasi-ordering and $\reflebsp$}\label{section:wqo}

A class $\cl{S}$ is \emph{well-quasi-ordered} (w.q.o.)  under a
pre-order $\preceq$ on $\cl{S}$, if for every infinite sequence
$\mf{A}_1, \mf{A}_2, \ldots$ of structures of $\cl{S}$, there exists
$i < j$ such that $\mf{A}_i \preceq \mf{A}_j$.  \emph{A priori}, there
is no reason to expect any relation between well-quasi-ordering and
$\reflebsp$. Surprisingly, the following holds.

\begin{theorem}[Thm. 11.2.2, Prop. 11.2.4, Chp. 11]\label{theorem:wqo-implies-lebsp}
Let $\cl{S}$ be a class of structures and $\cl{S}_k$ be as in
Proposition~\ref{prop:L-good-tree-reps-and-ebsp}.  If $\cl{S}_k$ is
w.q.o. under isomorphic embedding, then $\lebsp{\cl{S}, k}$ holds.
The witness function is not computable in general. Also, the converse
is not true in general.
\end{theorem}

While Theorem~\ref{theorem:wqo-implies-lebsp} channelizes the ongoing
research in w.q.o. theory~\cite{wqo-vs-clique-width} to our
model-theoretic studies (see Section~\ref{section:overview}), it also
gives a technique to show $\lebsp{\cdot, \cdot}$ for a class of
structures. For instance, while the fact that for $n > 1$, the class
$\cl{S}$ of $n$-dimensional grid posets satisfies $\febsp{\cdot, k}$
follows from the FO-FVC property of Cartesian product and
Theorem~\ref{theorem:op-tree-languages}, nothing can be inferred about
$\mebsp{\cdot, k}$ since Cartesian product does not have the MSO-FVC
property. But by Theorem~\ref{theorem:wqo-implies-lebsp},
$\mebsp{\cl{S}, k}$ is true since linear orders are w.q.o. under
embedding and hence so is their $n$-fold Cartesian product (which
gives $\cl{S}$). However, as MSO-SAT is undecidable over even
2-dimensional grid posets, any witness function for $\mebsp{\cl{S},
  k}$ is necessarily uncomputable.

%% file: results-obtained-post-thesis-3.0.tex
\section{Extensions of Dissertation Results} \label{section:post-thesis}

In this section, we let $\mathbb{N}$ denote positive integers. A
function $f: \mathbb{N} \rightarrow \mathbb{N}$ is called a
\emph{scale function} if it is strictly increasing. The
\emph{$i^{\text{th}}$ scale}, denoted $\scale{i}_{f}$, is defined as
the interval $[1, f(1)] = \{ j \mid 1 \leq j \leq f(1)\}$ if $i = 1$,
and $\left[f(i-1) + 1, f(i)\right] = \{ j \mid f(i-1) + 1 \leq j \leq
f(i)\}$ if $i > 1$. As mentioned in Section~\ref{section:overview},
various observations about the proof of
Theorem~\ref{theorem:abstract-tree-theorem} inspire the following
definition.

\begin{defn}[Logical fractal\footnote{A research proposal on this notion, written jointly with Anuj Dawar, has been granted a 3 year funding by the Leverhulme Trust, UK. The author has joined the University of Cambridge, UK as a post-doctoral research associate to pursue this research.}]\label{defn:logical-fractal}
  Given a class $\mc{S}$ of structures, $k \in \mathbb{N} \cup \{0\}$,
  and a pre-order $\preceq$ on $\cl{S}_k$ (where $\cl{S}_k$ is as in
  Proposition~\ref{prop:L-good-tree-reps-and-ebsp}), we say $\cl{S}$
  is an \emph{$(\mc{L}, k)$-fractal under $\preceq$}, if there exists
  a function $\vartheta_k: \mathbb{N}^2 \rightarrow \mathbb{N}$ such
  that (i) $\vartheta_k(n, \cdot)$ is a scale function for all $n \in
  \mathbb{N}$, and (ii) for each $m \in \mathbb{N}$ and each structure
  $\mf{A}$ of $\cl{S}_k$, if $f$ is the function $\vartheta_k(m,
  \cdot)$ and $|\mf{A}| \in \scale{i}_{f}$ for $i \ge 2$, then for all
  $j \neq i$, there exists a structure $\mf{B}$ in $\cl{S}_k$ such
  that (i) $\mf{B} \preceq \mf{A}$ if $j \leq i$, else $\mf{A} \preceq
  \mf{B}$, (ii) $|\mf{B}| \in \scale{j}_{f}$, and (iii) $\mf{B}
  \lequiv{m} \mf{A}$. We say $\vartheta_k$ is a \emph{witness} to the
  $(\mc{L}, k)$-fractal property of $\cl{S}$.
\end{defn}

Observe that if $\cl{S}$ is an $(\mc{L}, k)$-fractal under isomorphic
embedding, then $\lebsp{\cl{S}, k}$ is
true. Table~\ref{table:logical-fractals} lists a wide spectrum of
classes of computer science interest, that satisfy
Definition~\ref{defn:logical-fractal}. The listing in
Table~\ref{table:logical-fractals} is according to the complexity of
the relation $\preceq$ appearing in column 4. We put the function
$\nu(m) = \vartheta(m, 1)$ in column 5; we call this a
\emph{supporting function}. In all the cases listed where this
function is computable, the witness function $\vartheta(m, n)$ turns
out to be $O(\nu(m) \cdot n)$.  The results of
Table~\ref{table:logical-fractals} follow from the generalizations of
Proposition~\ref{prop:L-good-tree-reps-and-ebsp} and
Theorem~\ref{theorem:wqo-implies-lebsp} presented below, that are
established exactly like the latter. Towards these results, we first
relax the conditions of the abstract framework in part A of
Section~\ref{section:abstract-framework} to consider $\preceq$ instead
of $\hookrightarrow$.  Define an $(\mc{L}, \preceq)$-good tree
representation as an $\mc{L}$-good tree representation whose
operations satisfy the monotonicity properties of
Section~\ref{section:abstract-framework} where $\hookrightarrow$ is
replaced with $\preceq$. Call the associated representation map as
$(\mc{L}, \preceq)$-good.  We now consider special kinds of $(\mc{L},
\preceq)$-good representation maps $\mathsf{Str}$, those we call
\emph{$(\mc{L}, \preceq)$-great}, that satisfy the following
conditions: (i) $\mathsf{Str}$ is computable, and (ii) there is a
strictly increasing function $\beta: \mathbb{N} \rightarrow
\mathbb{N}$ such that for every $\tree{t}, \tree{s}$ in the domain of
$\mathsf{Str}$, if $\text{abs}(|\tree{t}| - |\tree{s}|) \leq n$, then
$\text{abs}(|\str{\tree{t}}| - |\str{\tree{s}}|) \leq \beta(n)$, where
$\text{abs}(\cdot)$ denotes ``absolute value''. We now present our
results which include an algorithmic metatheorem obtained by the
simple observation that the labelings, ``graftings'' and ``prunings''
of tree representations $\tree{t}$ as described in the proof sketch of
Theorem~\ref{theorem:abstract-tree-theorem}, which enable getting
``downward self-similarity'', are doable in time linear in
$|\tree{t}|$. (This is demonstrated in detail in the proof of
Proposition 3.2 in~\cite{abhisekh-csl17}.) Further, the same ideas in
a ``reverse direction'' give us ``upward self-similarity'' again in
linear time.  All the classes in Table~\ref{table:logical-fractals}
that have computable supporting functions admit $(\mc{L},
\preceq)$-great representations, whereby they are $(\mc{L},
k)$-fractals under $\preceq$, for the $k$ and $\preceq$ mentioned
against them.

\begin{theorem}\label{theorem:fractal-generator}
Given a class $\cl{S}$ of structures, let $\cl{S}_k$ be as before, the
class of structures obtained by labeling (possibly partially) the
elements of the structures of $\cl{S}$ with labels from $\{1, \ldots,
k\}$. For a pre-order $\preceq$ on $\cl{S}_k$, suppose $\cl{S}_k$
admits $(\mc{L}, \preceq)$-great tree representations. Then the
following hold:

\begin{enumerate}[nosep]
  \item $\cl{S}$ is an $(\mc{L}, k)$-fractal under $\preceq$ having a
    computable witness function $\vartheta_k: \mathbb{N}^2 \rightarrow
    \mathbb{N}$.
  \item There exists an FPT algorithm \textsf{Fractal-generator},
    parameterized by $m$ (the ``degree of logical self-similarity'')
    that, given a structure $\mf{A} \in \cl{S}_k$, an $(\mc{L},
    \preceq)$-great tree representation $\tree{t}$ of $\mf{A}$, and a
    number $j \ge 1$, outputs in time $g(m) \cdot (|\tree{t}| + j)$
    for some computable function $g: \mathbb{N} \rightarrow
    \mathbb{N}$, a structure $\mf{B} \in \cl{S}_k$, such that if $f =
    \vartheta_k(m, \cdot)$ and $|\mf{A}| \in \scale{i}_f$, then (i)
    $|\mf{B}| \in \scale{j}_{f}$, (ii) $\mf{B} \preceq \mf{A}$ if $j
    \leq i$, else $\mf{A} \preceq \mf{B}$, and (iii) $\mf{B}
    \lequiv{m} \mf{A}$. In short, \textsf{Fractal-generator} produces
    a logically self-similar structure at any given scale in FPT
    linear time.
\end{enumerate}
\end{theorem}

\input{fractal-table}

\begin{remark} 
Given that the ideas used in proving the ``existential''
Theorem~\ref{theorem:abstract-tree-theorem} are used to show the
``algorithmic'' Theorem~\ref{theorem:fractal-generator}, an important
question that arises is: how does one \emph{algorithmically construct}
the composition functions $f_{m, \mathsf{O}}$ for the $(\mc{L},
\preceq)$-great operations $\mathsf{O}$ that build the structures of
$\cl{S}_k$? This is necessary to get the function $g$ in
Theorem~\ref{theorem:fractal-generator} to be computable.  Here is how
we do it. We first observe that simply the \emph{existence} of an
$(\mc{L}, \preceq)$-great tree representation for $\cl{S}_k$ entails
the computable small model property for $\mc{L}$, and hence the
decidability of $\mc{L}$-SAT, over $\cl{S}_k$. We use this fact to
construct the set $\mc{E}$ of $\mc{L}[m]$ sentences corresponding to
those equivalence classes of the $\lequiv{m}$ relation, that have a
non-empty intersection with $\cl{S}_k$. For $\delta_1, \ldots,
\delta_r \in \mc{E}$, we find a model $\mf{A}_i \in \cl{S}_k$ for each
$\delta_i$ -- this is possible due to the small model property of
$\mc{L}$ over $\cl{S}_k$. Then $f_{m, \mathsf{O}}(\delta_1, \ldots,
\delta_n)$ is the $\lequiv{m}$ class of $\mathsf{O}(\mf{A}_1, \ldots,
\mf{A}_n)$. This argument shows that simply the \emph{existence} of
the FVC property for a (computable) operation, entails an
\emph{effective version} of this property. Finally, the generalization
below of Theorem~\ref{theorem:wqo-implies-lebsp}, along with
Theorem~\ref{theorem:fractal-generator}, explains all of
Table~\ref{table:logical-fractals}.
\end{remark}

\begin{theorem}\label{theorem:wqo-implies-logical-fractal}
Let $\cl{S}$ be a class of structures and $\cl{S}_k$ be as in
Theorem~\ref{theorem:fractal-generator}. For a pre-order $\preceq$ on
$\cl{S}_k$, if $\cl{S}_k$ is w.q.o. under $\preceq$, then $\cl{S}$ is
an $(\mc{L}, k)$-fractal under $\preceq$. The supporting function is
not computable in general. Also, the converse is not true in general.
\end{theorem}

%% file: fractal-table.tex
\newcommand*{\dittoclosing}{---''---}
\noindent
\begin{table}[h!]
\centering
\begin{footnotesize}
\begin{tabular}{|p{12pt}|p{2.25in}|p{0.9in}|p{1.12in}|p{0.6in}|}
\hline \centering No. & \centering Class & \centering $(\mc{L}, k)$-fractal for $(\mc{L}, k) = $  & \centering Fractal under $\preceq$ for $\preceq\,=$ & \centering Supporting function \tabularnewline
\hline
\multicolumn{5}{|c|}{\textsc{Posets}}\\
\hline
\multirow{2}{*}{}$1.$ & Regular languages of words/nested words & MSO, all $k$ & subword/sub-nested-word & non-elem.\\
$2.$ & Regular languages of trees (ordered, unordered, ranked, partially ranked) & MSO, all $k$ & subtree & non-elem.\\
$3.$ & Regular languages of traces & MSO, all $k$ & subtrace & non-elem.\\
4a. & $r$-dimensional grid posets & FO, all $k$ & subgrid & non-elem.\\
4b. & ~~~~~~~ \dittoclosing & CMSO, all $k$ & ~\dittoclosing& uncomp. \tabularnewline
$5.$ & All grid posets & FO, $k = 0, 1$ & subgrid & non-elem.\\
\hline
\multicolumn{5}{|c|}{\textsc{Graphs}}\\
\hline
$6.$ & Hamming graphs of the $n$-clique & FO, $k = 0$ & ind. subgraph & non-elem.\\
7a. & Disjoint unions of paths & FO, $k = 0, 1$ & ind. subgraph & exp. \\ 
7b. & ~~~~~~~ \dittoclosing & CMSO, $k = 0, 1$ &~ \dittoclosing & non-elem.\\
$8.$ & $d$-regular trees & CMSO, $k = 0$ & ind. subgraph & non-elem.\\
$9.$ & Graphs of tree-depth $\leq d$ & CMSO, all $k$ & ind. subgraph & $d$-fold exp.\\
$10.$ & Hereditary graph classes of shrub-depth $\leq d$ & CMSO, all $k$ & ind. subgraph & $d$-fold exp.\\
$11.$ & $m$-partite cographs (includes  threshold graphs, Turan graphs, cographs, $m$-partite threshold graphs)  & CMSO, all $k$ & ind. subgraph & non-elem.\\
$12.$ & Power graphs~\cite{wqo-vs-clique-width}$^1$ & CMSO, $k = 0$ & ind. subgraph & uncomp.\\
\multirow{3}{*}{}13. & Colored forests of height $\leq d$ & CMSO, all $k$ & surj. hom. img.;  & $d$-fold exp.\\
& & & vertex-minor; minor  & \\
$14.$ & Colored forests & CMSO, all $k$ & hom. img.; minor; vertex-minor & non-elem.\\
$15.$ & Graphs excluding a top. minor isomorphic to $P_k$ with each
edge duplicated & CMSO, $k = 0$ & topological minor & uncomp.\\
$16.$ & Series-parallel graphs$^2$ & CMSO, $k = 0$ & minor & comp. (?)\\
$17.$ & Graphs of tree-width$^2$ $\leq n$ & CMSO, $k = 0$ & minor & comp. (?)\\
$18.$ & Graphs of clique-width/NLC-width $\leq n$ & CMSO, all $k$ & a comp. reln.  & non-elem.\\
19a. & Graphs of rank-width$^3$ $\leq n$ & CMSO, all $k$ & vertex-minor & comp. (?) \\
19b. & ~~~~~~~ \dittoclosing & ~~\dittoclosing& a comp. reln. & non-elem.\\
$20.$ & All finite graphs & CMSO, $k = 0$ & minor; weak imm. & uncomp.\\
\hline \noalign{\smallskip}
\end{tabular}
\end{footnotesize}
\caption{A list of 20 logical fractals}
\label{table:logical-fractals}

\begin{flushleft}
\begin{footnotesize}
\begin{tabular}{ll}
$\dagger$: & \hspace{-8pt}ind.= induced; surj.= surjective; hom.= homomorphic;
  img.= image; reln.= relation; imm.= immersion \\
$\ddagger$: &\hspace{-8pt}comp.= computable; elem.=
  elementary; non-elem.= computable \& not elem.; exp.= exponential\\
1: & \hspace{-8pt}Anuj Dawar and the author have recently shown these graphs to admit MSO interpretability of grids.\\
2: & \hspace{-8pt}It is quite possible that these graphs have computable (but non-elementary) supporting functions.\\
3: & \hspace{-8pt}It is open whether these graphs have computably bounded $\lequiv{m}$-equivalent vertex-minors.\\
\end{tabular}
\end{footnotesize}
\end{flushleft}

\end{table}

%% file: conclusion.tex
\section{Conclusion}\label{section:conclusion}
The dissertation~\cite{abhisekh-thesis} introduces new dual
parameterized preservation properties that generalize the well-studied
notions of preservation under substructures (hereditariness) and
preservation under extensions; we call these \emph{preservation under
  substructures modulo $k$-cruxes} and \emph{preservation under
  $k$-ary covered extensions} respectively. These properties are
syntactically characterized in terms of $\Sigma^0_2$ and $\Pi^0_2$
sentences that have $k$ quantifiers in their leading block. This gives
a parameterized generalization of the classical {\lt} preservation
theorem, abbreviated $\glt{k}$, and also finer characterizations of
the $\Sigma^0_2$ and $\Pi^0_2$ classes than those in the
literature. We establish $\glt{k}$ and its variants first over
arbitrary (finite or infinite) structures and then over various
classes of finite structures, particularly those that are
\emph{dense}; these include several kinds of posets and subclasses of
graphs of bounded clique-width of active current interest. We show
that all of these classes are model-theoretically very well behaved:
they not only satisfy $\glt{k}$, but also the homomorphism
preservation theorem and a generalization of it akin to $\glt{k}$, and
a finitary analogue of the {\dls} theorem, all of these in effective
form and for MSO. These results come about by making the key
observation that the mentioned classes can be constructed using
operations that satisfy the Feferman-Vaught composition (FVC)
property. Extending these results (post thesis submission), we show
that the FVC property entails an effective finitary adaptation of the
full L\"owenheim-Skolem property (upward and downward), that we call
the \emph{logical fractal} property, and also entails linear time
algorithmic meta-theorems for CMSO. The above results collectively
seem to suggest that the FVC property might be playing a similar role
over the mentioned dense structures, as FO locality does over sparse
structures. We also show in the dissertation, a new connection between
well-quasi-ordering and model theory, that yields us another important
collection of dense classes that are model-theoretically
well-behaved. In summary, the thesis contributes new results to the
classical model theory literature, and also contributes to the
research programme of recovering classical model theory results over
finite structures. (A number of future directions, including a
conjecture, are presented in Chapter 12. Finally, a summary of the
thesis contributions appears in Chapter 13.)

%% file: appendix.tex
\appendix
\section{Relevance of Introduced Notions to Computer Science}\label{section:relevance}\label{appendix}

\input{relevance-of-notions-2.0}

%% file: relevance-of-notions-2.0.tex
\tbf{1. Fixed parameter tractability:} Table~\ref{table:psc-k-and-FPT}
lists 22 well-studied parameterized
problems~\cite{parameterized-algos} that are $PSC(k)$ for some $k$,
for $k$ related to the parameter of the problem\footnote{The author
acknowledges Rian Neogi, a Ph.D. student at the Institute of
Mathematical Sciences Chennai, in helping him prepare
Table~\ref{table:psc-k-and-FPT}.}.  For each problem below, if it is
$PSC(k)$, then it is not $PSC(k-1)$; also, the problem is expressible
using an $\exists^k \forall^*$ sentence.

\begin{table}[h!]
\begin{center}
\begin{tabular}{|m{0.5in}|m{2.8in}|m{0.8in}|}
  \hline
  \centering Sr. No. & \centering Problem parameterized by $k$ & \centering is \tabularnewline \hline
  \centering 1. & \textsc{Bipartite Matching} & $PSC(2k)$\\
  \centering 2. & \textsc{Multicolored Clique/Ind. Set} & $PSC(k)$\\
  \centering 3. & \textsc{Colorful Graph Motif} & $PSC( k)$\\
  \centering 4. &\textsc{Perfect Code} & $PSC(k)$\\
  \centering 5. &\textsc{Dominating Set} & $PSC(k)$\\
  \centering 6. &\textsc{Pseudo Achromatic Number} & $PSC(2\binom{k}{2})$\\
  \centering 7. &\textsc{Hitting Set/$d$-Hitting Set} & $PSC(k)$\\ 
  \centering 8. &\textsc{Ramsey} & $PSC(k)$\\
  \centering 9. &\textsc{Independent Set/Clique} & $PSC(k)$\\
  \centering 10. &\textsc{Set Cover} & $PSC(k)$\\
  \centering 11. &\textsc{Longest Cycle/Directed Cycle} & $PSC(k)$\\
  \centering 12. &\textsc{Set Packing/$d$-Set Packing} & $PSC(k)$\\
  \centering 13. &\textsc{Longest Path/Induced Path} & $PSC(k)$\\
  \centering 14. &\textsc{Subgraph Isomorphism} & $PSC(k)$\\
  \centering 15. &\textsc{MaxCut} & $PSC(2k)$\\
  \centering 16. &\textsc{Subset Sum} & $PSC(k)$\\  
  \centering 17. &\textsc{Maximum Matching} & $PSC(2k)$\\  
  \centering 18. &\textsc{Triangle Packing} & $PSC(k)$\\
  \centering 19. &\textsc{Max-SAT/Max-}$r$\textsc{-SAT} & $PSC(k)$\\
  \centering 20. &\textsc{Vertex Multiway Cut} & $PSC(k)$\\    
  \centering 21. &\textsc{Multicolored Biclique} & $PSC(2k)$\\
  \centering 22. &\textsc{Unique Hitting Set} & $PSC(k)$\\
  \hline
\end{tabular}
\caption{A list of 22 parameterized problems that are $PSC(k)$ for some $k$}
\label{table:psc-k-and-FPT}
  \end{center}
\end{table}

There are other FPT problems that can be readily seen to be $PSC(k)$
(and definable using $\exists^k \forall^*$ sentences) for some $k >
0$, such as \textsc{Chordal Completion, Feedback Vertex Set, Odd Cycle
Transversal} and \textsc{Vertex Cover}, but actually turn out to be
hereditary. The reason is that they talk of graphs that are $\leq k$
vertex deletions/edge modifications away from a hereditary property
(that is related to the example). We leave out these examples in
Table~\ref{table:psc-k-and-FPT} to demonstrate that $PSC(k)$ strictly
extends the scope and usefulness of hereditariness, and in an
interesting way.\footnote{\noindent The notion of $PSC(k)$ was first
formulated in the author's Master's thesis at IIT
Bombay~\cite{abhisekh-DDP}. The motivation was to develop methods for
TRDDC (Tata Research Development and Design Center) to formally verify
a software being developed for an insurance company in Pune, India.}

\vspace{3pt} \noindent \tbf{2. Finite model theory:}  A
set $X$ of vertices in a graph $G$ is said to be \emph{$d$-scattered}
in $G$ if for any two distinct vertices $u, v \in X$, their
$d$-neighborhoods in $G$ are disjoint. This notion is central to the
locality of FO: any FO sentence is equivalent to a Boolean combination
of ``local sentences'', where a local sentence asserts, for some $d,
m \in \mathbb{N}$, the existence of a $d$-scattered set of size $m$
satisfying some FO condition on the $d$-neighborhoods of the points in
the scattered set~\cite{libkin}. The notion of a scattered set appears
again in the definition of quasi-wide classes that were first
introduced in the context of the homomorphism preservation theorem in
the finite~\cite{dawar-quasi-wide}.  If $\mc{P}(d, m, r, N)$ is a
property of graphs asserting ``If the graph is of size $\ge N$, then
there exists a $d$-scattered set of size $m$ upon removal of $\leq r$
vertices'', then a graph class is \emph{quasi-wide} if there exists
$f:\mathbb{N} \rightarrow \mathbb{N}$ such that for every $d, m$,
there exists $N$ such that $\mc{P}(d, m, f(d), N)$ is true.  We now
observe that $\mc{P}(d, m, 0, N)$ is indeed $PSC(m)$ (and not $PSC(m -
1)$) for every $d, m, N \in \mathbb{N}$. And since $\mc{P}(d, m, r,
N)$ is $\leq r$ vertex deletions ``away from'' $\mc{P}(d, m, 0, N)$,
we get (by a similar reasoning as in the previous point) that
$\mc{P}(d, m, r, N)$ is also $PSC(m)$ (and not $PSC(m - 1)$) for every
$d, m, r, N \in \mathbb{N}$. We also observe that $\mc{P}(d, m, r, N)$
is (readily) expressible using an $\exists^{m+r} \forall^*$ sentence.

\vspace{3pt}
\noindent \tbf{3. Structural graph theory of sparse graph classes:} On
the dual front, $k$-ary covers play a central role in graphs of
bounded expansion and nowhere dense graphs~\cite{nesetril-sparsity},
as seen from the characterizations of these graphs, stated below:

A class $\mc{C}$ of graphs has \emph{bounded expansion}
(is \emph{nowhere dense}) if, and only if, there exists
$f:\mathbb{N} \rightarrow \mathbb{N}$ such that for every integer $k$
(for every integer $k$ and every $\epsilon > 0$), every graph
$G \in \mc{C}$ (every graph $G \in \mc{C}$ of order $n \ge
f(k, \epsilon)$) has a \emph{$k$-ary cover} $R$ consisting of graphs
of tree-depth at most $k$, where every vertex of $G$ is in $\leq f(k)$
(in $\leq n^\epsilon$) structures of $R$.

Very recently~\cite{FO-interpretation-bounded-expansion}, graph
classes that have \emph{structurally bounded expansion} have been
introduced in the context of investigating dense structures for
algorithmic metatheorems. These are graph classes that are obtained
from bounded expansion classes by means of first-order
interpretations. It turns out that these classes also have a
characterization in terms of $k$-ary covers, as stated below:

A class $\mc{C}$ of graphs has \emph{structurally bounded expansion}
if, and only if, there exist functions $f,
g:\mathbb{N} \rightarrow \mathbb{N}$ such that for every integer $k$,
every graph $G \in \mc{C}$ has a \emph{$k$-ary cover} $R$ consisting
of graphs from a graph class of shrub-depth at most $g(k)$, where
every vertex of $G$ is in $\leq f(k)$ structures of $R$.